\documentclass[preprint]{revtex4-1} 

\usepackage{amsfonts} 
\usepackage{graphicx}
\usepackage{amsmath, amssymb}
\usepackage{slashed}
\usepackage{braket}
\usepackage{physics}
\usepackage[bottom]{footmisc}
\graphicspath{  {images/} }
\begin{document}


\title{A microscopic model of wave-function dephasing and decoherence in the double-slit experiment}

\author{Satish Ramakrishna}
\affiliation{Department of Physics \& Astronomy, \\
Rutgers, the State University of New Jersey, \\
136 Frelinghuysen Road
Piscataway, NJ 08854-8019}
\email{ramakrishna@physics.rutgers.edu} 
\begin{abstract}

\section{Abstract}

The act of measurement on a quantum state is supposed to  ``dephase'', then  ``decohere'' and ``collapse'' (or more precisely  ``register'' and ``reduce'') the state into one of  several eigenstates of the operator corresponding to the observable being measured. This measurement process is sometimes described as outside standard quantum-mechanical evolution and not calculable from Schrödinger's equation. Progress has, however, been made in studying this problem with two main calculation tools - one uses a time-independent Hamiltonian, while a rather more general approach proving that decoherence occurs under some generic conditions.

The two general approaches to the study of wave-function collapse are as follows. The first approach, called the ``consistent'' or ``decoherent''' histories approach, studies microscopic histories that diverge probabilistically and explains collapse as an event in our particular history. The other, referred to as the ``environmental decoherence'' approach studies the effect of the environment upon the quantum system, to explain wave-function decoherence which is produced by irreversible effects of various sorts. However, as we know, wave-function collapse is not related to thermal connection with the environment, rather, it is inherent to how measurements are performed by macroscopic apparata.

In the ``environmental decoherence'' approach, one studies decoherence using a Markovian-approximated Master equation to study the time-evolution of the reduced density matrix (post dephasing) and obtains the long-time dependence of the off-diagonal elements of this matrix.

The calculation in this paper studies the evolution of a quantum system starting with ``dephasing'' followed by the effects of the environment with some differences from prior analyses. We start from the Schrödinger equation for the state of the system, with a time-dependent Hamiltonian that reflects the actual microscopic interactions that are occurring. Then we systematically solve (exactly) for the time-evolved state, without invoking a Markovian approximation when writing out the effective time-evolution equation, i.e., keeping the evolution unitary until the end. This approach is useful, and it shows that the system wave-function will explicitly ``un-collapse'' if the measurement apparatus is sufficiently small. However, in the limit of a macroscopic system, this ``dephasing'' quickly leads to ``decoherence'' - collapse is a temporary state that will simply take extremely long (of the order of multiple universe lifetimes) to reverse. This has been attempted previously  and our calculation is particularly simple and calculable. We make some connections to the work by Linden {\it et al} while doing so.

The calculation in this paper has interesting implications for the interpretation of the Wigner's friend experiment, as well as the Mott experiment, which is explored in Sections V and VI (especially the enumerated points in Section VI). The upshot is that as long as Wigner's friend is macroscopically large (or uses a macroscopically large measuring instrument), no one needs to worry that Wigner would see something different from his friend. Indeed, Wigner's friend does not even need to be conscious during the measurement that she conducts. It also allows one to reasonably interpret some of the more recent thought experiments proposed.

In particular, as a result of the mathematical analysis, the short-time behavior of a collapsing system, at least the one considered in this paper, is not exponential. Instead, it is the usual Fermi-golden rule result. The long-term behavior is, of course, still exponential. This is a second novel feature of the paper - we connect the short-term Fermi-golden rule (quadratic-in-time behavior) transition probability to the exponential long-time behavior of a collapsing wave-function in {\it {one}} continuous mathematical formulation.

\end{abstract}
\maketitle 

\section{Introduction}

The study of quantum measurement has had a long history\cite{Balian6,Balian5,Borg,Balian1,Popescu,Everett,Zeh,Zwanzig,Vedral}, beginning with the earliest discoveries in quantum mechanics. The studies of dephasing and decoherence in quantum computers represent practical realizations of the measurement problem, as wave-functions representing quantum states of computers slowly transition from superposed to mixed states due to interactions with the environment. 

The process of measurement is expected to consist of three separate phases - dephasing of several components of the wave-function due to interactions with the measuring device, decoherence due to dissemination of the initial phase information amongst many degrees of freedom in a bath, which directly then leads to collapse of the wave-function into one of the many possible measurement outcomes.
There have been two broad approaches to describe this process of wave-function collapse from standard quantum mechanics. 

The first is a variety of approaches that can be collectively referred to as the ``decoherent histories'' method. These were championed by Griffiths, Omnes etc. \cite{Griffiths, Omnes, Hartle1, Hartle2} and posit, akin to the Many Worlds hypothesis of Everett \cite{Everett}, consistent microscopic histories that diverge probabilistically according to the laws of quantum evolution, but stay internally consistent. They resolve the issue of "collapse" (if it is an issue at all) by stating that nothing has actually collapsed, we just sit on one possible branch of several histories. Essentially, it answers the question - ``Why did the measurement of a quantum system produce {\bf this} result?'', by asserting that ``every result was actually produced, but the other results were produced in other universes that are inaccessible to us''. While this may qualify in grammar as an answer, it is (in the opinion of this author) unsatisfying to introduce elements into our knowledge of the world that are unknowable in principle.

The second, usually referred to as the ``environmental decoherence'' method \cite{Maximilian}, was pioneered by Zeh e\cite{Zeh} and exemplified in work by Zurek \cite{Zurek} and in a different manner by Weisskopf \& Wigner \cite{Weisskopf} and Allahverdyan {\it et al \:}\cite{Balian1}. Here, one explicitly couples an external heat bath to the quantum system and studies a master (Liouville) equation for the density matrix with the system coupled to the heat bath. Dephasing, then decoherence is achieved by the exponential vanishing of off-diagonal elements of the density matrix (the overlap between different eigenstates of the underlying quantum system). The master-equation method is also used in quantum-trajectory and quantum-jump analyses \cite{Sorgel}, with the addition of intermediate measurements to condition the density matrix.

In this second approach, one uses a variety of assumptions to derive a tractable problem specified in terms of a reduced density matrix. Among the common methods to simplify the problem is a Markovian assumption, which allows the Lindblad operator to be simplified by the assumption of a separable (product) density matrix between system and environment\cite{Maximilian}.

A very similar problem has been studied intensively by Allahverdyan et. al\cite{Balian1, Balian2, Balian3,Balian4}, using the above density matrix formalism and a time-independent Hamiltonian. In these, the authors study a spin system coupled to a large magnetic system in a metastable state, near the Curie-Weiss transition and further coupled to a heat bath. They demonstrate that the spin-system's density matrix, when traced over its environment, moves from possessing off-diagonal elements to one with purely diagonal elements, consistent with a mixed (``decohered'') state. The decoherence happens very quickly, at a rate proportional to the (square root of the) macroscopic number of degrees of freedom in the magnet that undergoes the transition (i.e., the measuring device). However, the ``registration'' of this ``decoherence'', {\it i.e.,} collapse, at the instrument pointer, is found to be much slower, driven by the relaxation of the diagonal elements in the density matrix. The irreversibility of the measurement process is produced by an explicit macroscopic (Curie-Weiss) phase-transition that also ends up storing the result. A detailed analysis of this, as well as a clear definition of terms is included in \cite{Balian5}.

Why are we studying the same problem, with a time-dependent Hamiltonian?

We do so in order to address some of our concerns with the approach detailed above. A key determinant of the calculation in Allahverdyan {\it et al} is that the interaction between the quantum system and the measurement apparatus is directly proportional to the total number of degrees of freedom ($N$) in the apparatus. This is usually not true in a real experiment. In fact, when we make a ``measurement'' at a double slit\cite{Feynman}, the electron interacts with {\bf one} photon. {\it {The key is not how many particles interact with the quantum system initially - the key is how the measurement apparatus distributes the effect of the interaction amongst a macroscopic number of internal degrees of freedom in a manner that is extremely difficult to reverse}}. We make this precise later (see Section IV). In fact, much as occurs in thermalization, if there are several ways to distribute energy over a large number of individual eigenstates, the recurrence rate of an ``ordered'' state is related to the fraction of ordered states over all possible states $N$. This is not just of $\sim \frac{1}{N}$, it is of $\sim e^{-N}$, which accounts for why measurements made in most laboratories never un-collapse ({\it i.e.,} the same measurement manifests itself as different values for different observers or correctly functioning instruments). In a sense, we are analyzing the ``dephasing'' process in detail through the microscopic interaction.

Once dephasing is achieved, decoherence just needs a large measuring apparatus and the process is akin to, but not exactly the same as approach to thermodynamic equilibrium \cite{Balian5}. Let's explain this, in our context, with an analogy. Consider $N$ independent, non-interacting hard sphere particles in a box. Let's divide the interior of the box into small cubes - $M$ of them; we also assume that $N$ and $M$ are large. If we define an ``ordered'' state as one where all the $N$ particles are sitting in the same cube, there are only $M$ such possibilities, versus $M^N$ possibilities for ``dis-ordered'' states. In the same way, as we will see later in our analysis, there are a multiplicity of ways of distributing the photons produced in the measuring apparatus among a large number of states. These are similar to the multiplicity of ``disordered'' states versus the rather small (one) number of a pristine state for the system in one of two possible states.

Additionally, while the present paper describes a study of the Schrödinger equation with a time-dependent Hamiltonian directly (hence along the lines of the ``environmental'' approach), it is different in that we do not require a Curie-Weiss type of paramagnetic-ferromagnetic transition to achieve similar results - while the similarity to a phase transition is striking as an explanation for why one measurement results rather than another, it naturally emerges from a first-principles unitary calculation in our approach. In addition, a central point in our thesis is that one needs to add energy into the system to make a measurement, it is not a conservative operation upon the measured system. However, it is similar in that the macroscopic number of variables that are affected are the root of the rapid decoherence. Collapse is then the result of adding variables that include the macroscopic measuring apparatus (and maybe one or more observers, though that is not a requirement) into the description.

The interesting off-shoot of our approach is that it is clear that the result of a measurement is random, but it is possible to reverse the result and revert to an unmeasured (``un-collapsed'') state as long as the measuring apparatus is small enough ({\it i.e.,} not macroscopic). Indeed, while the approach easily produces decoherence, introducing a macroscopic measuring apparatus into the calculation forces collapse for that measuring instrument. Furthermore, since ``undoing'', purely through quantum fluctuations, the modified state (post-measurement) of a macroscopic measuring apparatus (and needless to say our hapless experimental colleague) is so difficult and time-consuming that all future observers will agree with the result obtained by this apparatus and experimenter.

This calculation supports the viewpoint that the Observer as well as the Quantum System can be thought of as part of a larger quantum state, which hasn't really collapsed at all, simply moved around in Hilbert space in response to an interaction Hamiltonian. Thus, the consequences of the act of measurement can indeed be modeled by standard quantum mechanics, straightforwardly from Schrödinger's  equation with the addition of energy into the system at periods of interaction. No additional assumptions are needed to produce decoherence and wave-function collapse. The approach to treating the observer as part of the measurement process is quite well-known, researched intensively and described lucidly in \cite{Coleman}.

\section{Method}

To summarize the approach in a few sentences, which also connects this analysis to the Wigner's friend puzzle (see section at the end),
\begin{enumerate}
\item Every measurement apparatus can produce multiple possible outcomes as a result of interaction with a measured system.
\item At every instant of time, the underlying system is interacting with the measurement apparatus to produce every possible outcome as a result of the interaction Hamiltonian. This ``fluctuation'' (called ``dephasing'' in \cite{Balian5}), while similar in origin to a statistical fluctuation, occurs because the system being studied is interacting with several small particles in the measuring system, exchanging energy with them all the time. An analogy (made precise later) is to a large ({\it i.e.,} macroscopic)  number of harmonic oscillators interacting with each other. Though their motion is entirely deterministic, for all practical purposes, we could think of the energy variations of each oscillator as random statistical fluctuations. These interactions involve, and transfer energy between, varying numbers of oscillators/particles.
\item The number of particles involved in determining each of the outcomes $i_1, i_2, ...i_N$ is $M_1, M_2, ..., M_N$ at every instant.
\item Each outcome is the result of one of these interaction ``fluctuations'' - the time it would take for the outcome to be ``unmeasured", i.e., for the fluctuation to reverse direction, is of the order of $\sim e^{M_1}, e^{M_2}, ..., e^{M_N}$. This is discussed in Section IV (recurrence times).
\item The function $f(M)=e^M$ is such that if $M>N$, then $f(M)>>f(N)$ and $f'(M)>>f'(N)$.
\item While an interaction ``fluctuation" is reversing itself, there is a chance that more particles in the measuring system will become involved in that fluctuation. Hence a longer-lived fluctuation, given the previous clause, just wins eventually.
\item Any ``fluctuation", even the longest lived one, will eventually reverse itself. The time to reverse is larger, the larger the number of particles involved. For instance, an experimenter might make a measurement, write it down in her notebook, publish a paper, involving more and more particles, thereby making the particular fluctuation that produced one measurement that much more time-consuming to reverse. In every case, the time it would take to witness a reversal is of the order of $\sim e^M$, where $M$ is the number of particles involved. 
\item ``Fluctuations'' of this sort cause ``dephasing'' and eventually produce decoherence when large numbers of particles get involved. The ``fluctuation'' with the largest number of particles involved (including the experimental apparatus etc.) is the one the measurement ``collapses'' into. 
\end{enumerate}

The form of the paper is as follows. We study the classic double-slit experiment, with an idealized photomultiplier tube and add an important ingredient - measurement is a process where external energy is used to detect and amplify weak signals, which are then registered on a macroscopic device. This energy could come through the energy of a photon that scatters off an electron, or the subsequent photo-amplification in our process. Such amplification is necessary to prevent noise from being mistaken for a measurement.

After setting up the experiment and the equations for the time-evolution, we study the form of the time-evolution equation, recognize the form and propose a simpler version that can be numerically studied, as well as analytically investigated. We present the numerical results for a large number of subsequent states and then show analytically that both Fermi's Golden Rule and the exponential "collapse" into one state can be obtained from the same formalism. 

Then we make some connections to other approaches and studies - in particular, those of Linden {\it et al}, Wigner's friend and the Mott experiment. We also comment on the recurrence times for the collapse to reverse.

\section{The double-slit experiment}

\noindent Consider the description of the standard double-slit experiment. A source of electrons is on the left in Fig. 1. It shoots electrons, at an extremely slow rate, at an absorbing screen (on the right). Interposed between the screen and the electron gun is another absorbing screen, which has two slits (labeled $1$ and $2$). Between the slits and assumed to be aimed at slit $1$, there is an intense source of high-energy ``locator'' photons, that can scatter (if they interact) off the  electrons. Most of the ones that scatter off electrons are picked up by the input maw of a photo-multiplier tube. The energy of the photons that scatter off the electrons is $\omega_p$, while the energy of the electrons is $\omega_e$ (we use natural units throughout the calculation, so $\hbar = 1$).

\begin{figure}[h!]
\caption{Schematic of the Double-Slit experiment}
\centering
\includegraphics[scale=.75]{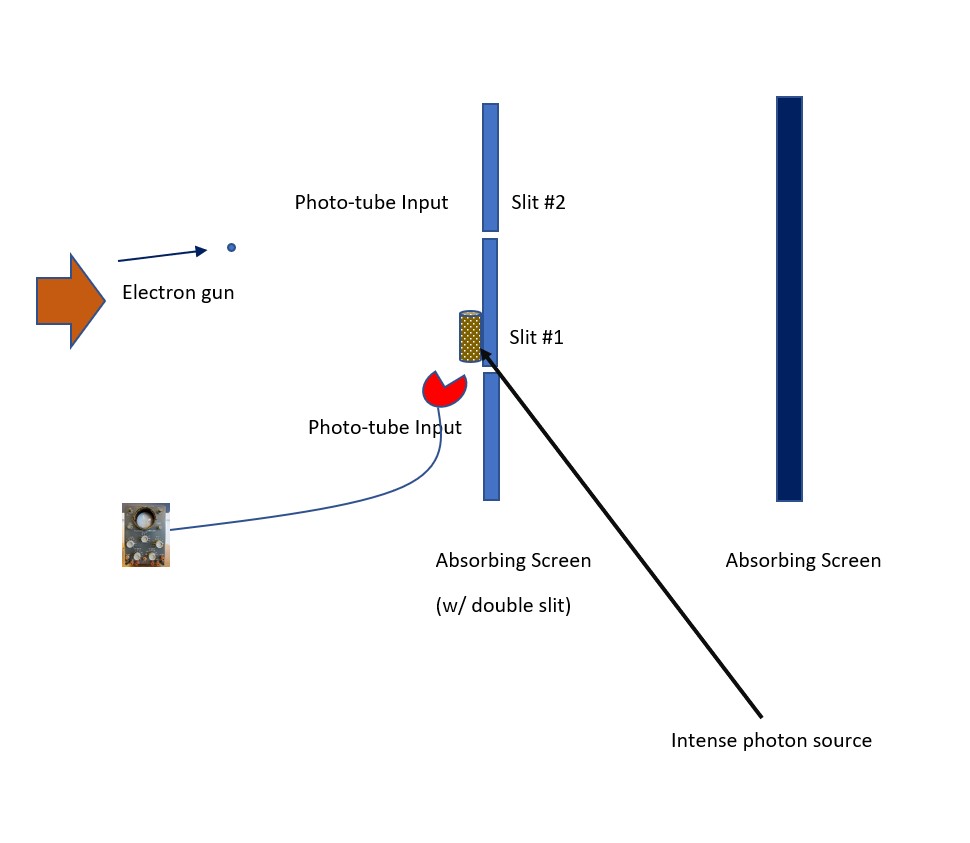}
\end{figure}

\noindent These ``locator'' photons are captured by an idealized photomultiplier tube after scattering. We don't need, for the purposes of the calculation, to understand the details of how the exact process of amplification occurs - we just assume that one photon, when absorbed, produces $N$ more photons and the process continues. We just need to realize that the end result is a bright spot, that emits several (a macroscopic number of) photons of several frequencies, from a point on a display tube. This point on the display tube is, therefore, directly connected to the reception of a single locator photon at the input end of the amplifier.

\noindent The physical description of the  Hamiltonian applicable to the problem is as follows. There is a ``free'' term that represents the free electron, with energy $\omega_e$. This electron could be at slit $1$ or at slit $2$, these are represented by subscripts on the number operator for the electrons. There is also a ``free'' term that represents the free photons, of frequency $\omega_p$ (for the locator photon that directly scatters off the electron at slit $1$), while $\omega_i$ represents the frequencies of the photons that are subsequently produced by the initially scattered photon (and thence) at the photomultiplier tube. We ignore photons from the intense source that do not scatter off the electron, since the photomultiplier input ignores the direct beam and only looks for scattered photons. We are modeling the process where the electron interacts with one photon at the slit, which, then at the photomultiplier tube, cascades, producing $N$ other photons in each state $i$ with each subsequent interaction. 

In a real photo-multiplier tube, the locator photons excite a few electrons, each of which then produces a few more electrons. The process cascades and a large current (millions of times larger than the initially excited current) is produced.  In our case, we assume the locator photons excite a few photons, which each excite a few more photons and so on, to produce a macroscopic number of photons in a continuum of photon energy states. This constitutes ``measurement'' by the device. We consider the initial wave-function of the electron to have collapsed into one that is definitely localized at slit $1$ when the amplitude for the purely superposed state falls very close to 0.

There is a subtle point to consider. We purposely select, out of all the possible ways to select independent vectors in the Hilbert space of the problem, vectors that correspond to our particular situation. In particular, one might ask why we assume that the electron has indeed gone through slit $1$ when we see a macroscopic signal from the photomultiplier tube at slit $1$. The answer is  that we ``define'' the situation as such - we declare the result of the measurement is that {\underline if} we see a macroscopic signal at slit $1$, it implies that the electron indeed went through slit $1$. Our observation that the interference pattern on the screen gets replaced by a ``clump'' pattern confirms that this assignment is right. To put it pithily, we have learnt that when we see a car headed towards us, it is useful to step out of the way. In the density-matrix formalism, this just means that we have zero off-diagonal elements post-measurement\cite{Balian2}.

In our approach below, we choose independent vectors in Hilbert space corresponding to what we define as the inference of a measurement. In addition, our choice of basis vectors in Hilbert space is determined by how we interpret the results of our measurements - our theory decides what we have observed.

We write the hermitian Hamiltonian as
\begin{eqnarray}
{\cal H} = {\cal H}_{electron} + {\cal H}_{photon} + {\cal H}_{int}^{detector} + {\cal H}_{int}^{PM \: initial}+ {\cal H}_{int}^{PM\: tube}
\end{eqnarray}
\noindent where each piece is explained below. First, we start with the Hamiltonian for the ``free'' electrons and photons. The ``free'' term for electrons counts the electrons at each slit ($1$ and $2$), while the ``free'' photon term counts the photons in various states in the photomultiplier tube.
\begin{eqnarray}
{\cal H}_{electron} = \omega_e \left( c_1^{\dagger} c_1 + c_2^{\dagger} c_2 \right) \nonumber \\
{\cal H}_{photon} = \omega_p \gamma_0^{\dagger} \gamma_0 + \sum_{i=1}^{R} \omega_i \gamma_i^{\dagger} \gamma_i 
\end{eqnarray}
\noindent Then we add interactions between the electron and the first locator photon (we assume the interaction parameter $\Gamma_1$ is real in this example). This interaction is modeled as between the electron at slit \# $1$ and the photon in mode $0$ - there is no photomultiplier at slit $2$, so no such interaction is possible there. In order to ensure energy conservation, energy needs to be supplied through the interaction. One may consider the interaction parameter as a background field, with a time dependence, which allows the photon to be produced in the presence of the electron at slit $1$ and emerge with energy $\omega_p$. In Appendix III, we describe how this interaction would come about from the usual QED vertex and a background field. 
\begin{eqnarray}
{\cal H}_{int}^{detection} = \Gamma_1 e^{-i \omega_p t} \gamma_0^{\dagger} c_1^{\dagger} c_1 + \Gamma_1  e^{i \omega_p t} \gamma_0 c_1^{\dagger} c_1
\end{eqnarray}
\noindent Next, we add the interactions that lead to the creation of child photons (in any of $R$ photon states) from this first locator photon at the photomultiplier tube ($N$ is a multiplication fraction, usually $2-3$). Again, energy is inserted into the system through the interaction vertex, the first ``enhancement'' being modeled by another auxiliary field with time-dependence. 
\begin{eqnarray}
{\cal H}_{int}^{PM \: initial} =\sum_{i=1}^R \bigg( \Gamma_2^i  e^{-i N \omega_i t + i \omega_p t}  (\gamma_i^{\dagger})^N \gamma_0 + \Gamma_2^i e^{i N \omega_i t - i \omega_p t}\gamma_0^{\dagger} \gamma_i^N \bigg)
\end{eqnarray}
\noindent Next, these photons create other photons in the available states, with  similar multiplication factors (we use $M_1$), again using the external measurement system to add energy. 
\begin{eqnarray}
{\cal H}_{int}^{PM \: tube} = \sum_{i=1, j=1; i<j; i \ne j}^R \bigg( s \: e^{i \omega_i t-i M_1 \omega_j t} (\gamma_j^{\dagger})^{M_1} \gamma_i + s \: e^{-i \omega_i t+i M_1 \omega_j t} \gamma_i^{\dagger} (\gamma_j)^{M_1} \bigg) 
\end{eqnarray}

This Hamiltonian is explicitly time-dependent  and energy is inserted from the outside through the interaction. Apart from a choice in how we display the results, these interactions are entirely plausible approximations to what actually happens in the real-world.

The scale of the couplings and photo-multiplication is, $\Gamma_2 \approx \Gamma_1 \approx  s$, while $N, M_1$ are of ${\mathcal O}(1-10)$. The number of modes $R$ of photons that are subsequently produced, is macroscopic and large. Note a crucial difference between this approach and that of Allahverdyan {\it et al} \cite{Balian1} - here the microscopic interaction is between few particles. It is the subsequent re-distribution into a macroscopic number of possible states that causes the dynamics to take the form that it does. In simple terms, it is not that a billiard ball is stopped by a wall - it is stopped by a small collection of smaller particles that re-distribute the energy amongst a host of other small particles that causes the billiard ball to stop.

As one can see, once a photon interacts with the electron, it subsequently produces more photons and the process cascades exponentially within the states available to the photons - subsequent processes distribute this energy into other states in the continuum.

\subsection{Framework of the Calculation}

How is the double-slit experiment understood?  In the classic experiment, where no intermediate measurement is done, we have the electron arrive at the screen in the superposed state
\begin{eqnarray}
\ket{\psi} = \frac{\ket{1} + \ket{2}}{\sqrt{2}} \nonumber
\end{eqnarray}
Represented in the position representation on the screen (assumed to be one-dimensional along the x-direction shown in Fig. 14 in Appendix I), this is
\begin{eqnarray}
\bra{x}\ket{\psi} = \frac{\bra{x}\ket{1} + \bra{x}\ket{2}}{\sqrt{2}} \nonumber
\end{eqnarray}
With this information, we can deduce (by the usual application of the Born rule), that the probability density $P(x)$ of locating an electron on the absorbing screen is, in the limit of looking far from the central region between the slits,
\begin{eqnarray}
P(x) = \frac{|\bra{x}\ket{1} + \bra{x}\ket{2}|^2}{2}= \frac{2}{x^4} \cos^2 \frac{k \: d \: x}{R} \nonumber
\end{eqnarray} 

When we interpose a photomultiplier tube at slit $1$, then the superposed state becomes (with the electron at slit $1$ or $2$ and the photomultiplier tube's photons in various occupation number states)
\begin{eqnarray}
\bra{x}\ket{\psi} = \frac{\bra{x}\sum_{n_0, ..., n_R}\ket{1; n_0, n_1, ..., n_R} + \bra{x}\sum_{n_0, ..., n_R}\ket{2; n_0, n_1, ..., n_R}}{\sqrt{2}} \nonumber
\end{eqnarray}
where $n_0, n_1, ..., n_R$ are the numbers of photons excited in the photomultiplier tube (located at slit $1$).

If the photomultiplier tube is off, we can safely assume that the only contribution to the sum comes from $n_0, ..., n_R=0$. In that case, we are reduced to the same calculation as we had above for the undisturbed double-slit experiment.

The action begins when the photomultiplier tube is turned on. The electron could not have excited any photons in the tube if it were at slit $2$. So the second term above has only one term in the sum over $n_i$, i.e., with $n_0, ..., n_R=0$. The first term is a different story, however. First, there is a strong chance that the electron slips through without being observed by the photomultiplier tube at all. In that case, there is a term in the sum corresponding to $n_0, ..., n_R=0$ - this is the case when no interaction has occured and the electron system is undisturbed.  However, there are other terms where the electron has scattered off the photomultiplier target and excited many photons (in the photomultiplier tube and beyond) into several states. As we will see in what follows, the result is a collection of different states - with many photons occupying a variety of different states within the photomultiplier tube, even in the experimenter's eyes, her iPad etc. The state now can be written as
\begin{eqnarray}
\bra{x}\ket{\psi} = \frac{1}{\sqrt{2}}  \bigg[ \bra{x}\bigg(\alpha \ket{1; n_0=0, n_1=0, ..., n_R=0}+ \beta \ket{1; n_0=1, n_1=0, ..., n_R=0} + ...\bigg) + \nonumber \\
\bra{x}\ket{2; n_0=0, n_1=0, ..., n_R=0} \bigg] \: \: \: \: \:  \: \: \: \: \:  \: \: \: \: \:  \: \: \: \: \:  \: \: \: \: \:  \: \: \: \: \:  \: \: \: \: \: \nonumber
\end{eqnarray}
where $\alpha$ represents the coefficient of the state with no interaction (initially $\alpha=1$), $\beta$ that of the state with one photon in state $n_0$ (initially $\beta=0$), etc. As we will see in what follows, if we set up the measuring apparatus with enough different niches ({\it i.e.,} states) for a macroscopic number of photons to occupy, $\alpha \rightarrow 0$ very rapidly and the other terms ($\beta$ etc.)  gain in value.

A remark is in order about the position space representation of these photon states, {\it viz.}, they are localized to within the photomultiplier tube and the experimenter's eyes and iPad. Importantly, they have phases that bear no relation to the phase of the slit $2$ term or other terms localized to slit $1$ (in fact, importantly, in the occupation number representation, the phase is indeterminate \cite{Murayama}). Hence, if we were to compute $|\bra{x}\ket{\psi}|^2$, apart from the usual interference term,  the loss of deterministic phase relationships (and the macroscopic number of terms) between the various other terms in the expression results in 
\begin{eqnarray}
P(x) = \frac{(\alpha+\alpha^*)}{2}|\bra{x}\ket{1} |\bra{x}\ket{2}| + \frac{1}{2}|\bra{x}\ket{2}|^2  + \frac{1}{2} (|\alpha|^2+|\beta|^2+...) |\bra{x}\ket{1}|^2 \nonumber
\end{eqnarray}
Of course, $|\alpha| \rightarrow 0$ very quickly once the measurement has occurred. In the situation, the remnant $(|\beta|^2+...) \rightarrow 1$ so that the appropriate amplitude is distributed amongst a large number of states of the measuring apparatus

Note that adding the experimental apparatus (as well as registers and pointers on the apparatus) into the set of final states indicates that together with decoherence (the final probability involves adding probabilities of each individual outcome), we have also achieved registration/collapse - the experimenter (or her apparatus) has perceived that the electron went through slit $1$ and this has destroyed the interference of the two paths. In this case, decoherence implies ``collapse'' because there was only one {\it yes/no} measurement involved. In general, one can only say that one of several outcomes has been selected, which one is impossible to predict for it depends on the precise microscopic interactions that occurred.

\subsection{``Collapse/Registration'' or ``Decoherence''}
This probabilistic outcome, i.e., with decoherence and subsequent registration, is consistent with \cite{Balian5}. $P(x)$ simply describes what ({\it i.e.,} the arriving electron density) an observer would see at the absorbing screen. The reason this observer would see a ``clump'' pattern is that the measuring apparatus at slit $1$ has picked up electrons that went through slit $1$. The ``other'' electrons (the ones that were not picked up by the measuring apparatus) would still show interference - for them there was no wave-function ``collapse''.

One could question whether this paper demonstrates ``collapse'' rather than just ``dephasing'' or `` decoherence''. Regarding this objection, surely it is unreasonable for this paper, at this time, to select a particular outcome for the double-slit experiment for a particular experiment. All one can demonstrate is that
\begin{enumerate}
\item there is a deterministic process with many ``participants'' (photons) that is akin to a random process by which an outcome is selected
\item there is a reason why one of those processes "wins", linked to the number of photons that populate the detector for that outcome
\item  that  "collapsed" outcome is the "winning" outcome
\item in the absence of detailed knowledge about the exact random processes and their state, all one can state is the probability of various outcomes and note that there is no entanglement between those outcomes anymore, since the phase information gets erased in the occupation number representation (see previous  note and the reference to Murayama \cite{Murayama})
\end{enumerate}

Surely, one cannot, writing this paper, decide what outcome is going to emerge! All one can do is to provide a detailed mechanism by which such an outcome emerges and demonstrate that it would be not appear deterministic at all.

Suppose the measuring apparatus were {\bf not} macroscopic. Even one photon interacting with the electron introduces a random phase term into the second term of the state above. This would, by the above argument, produce collapse. However, if we reversed the effect of the random phase term (as one does in the Delayed Choice Eraser experiments), it is very clear that one can produce ``un-collapse'' and recover interference patterns. The key is to control precisely (and keep small) the interactions the electron has with external systems so one can precisely reverse the ``random'' phase multiplier thus introduced.

If one asks the question why did the universe ({\it i.e., }the electron) pick one outcome over another, it is clear both outcomes (electron going through either slit $1$ or $2$) had an initial shot at success. The dynamics of a vast, macroscopic system of photons needed to be sparked off with one minor detection, after which the host of other degrees of freedom were affected. {\bf If} that initial detection happened, and was correctly amplified to affect a macroscopic number of degrees of freedom, the experimental apparatus and the experimenter included; indeed, for that experimenter (and all subsequent interactors), the wave-function has collapsed. Which of the several outcomes {\it wins} is a result of microscopic interactions that need to involve the most number of measuring apparatus particles.

{\it Dephasing followed by decoherence leads to collapse once we include the states of the measuring apparatus, into the quantum mechanical description. To do this, we need a properly defined interaction Hamiltonian between the system and the measuring apparatus.}

\subsection{The States}

We will now write down some key states in the Fock space of the system. The total state of the system is the weighted sum of all possible states - first, an `unperturbed'', i.e., {\it un-decohered and un-collapsed}, state and next, several other ``perturbed'', i.e., what we could refer to as  {\it decohered and collapsed}, states.

\noindent First, we represent the state of the unperturbed ({\it i.e.,} uncollapsed) system as
\begin{eqnarray}
\ket{\psi}_0 = \frac{a \ket{1}_e + u \ket{2}_e}{\sqrt{2}} \: \ket{n_0=0, n_1=0, n_2=0, ... , n_Q = 0}
\end{eqnarray}
The kets $\ket{1}_e, \ket{2}_e$ represent the state of the electron localized either to slit $1$ or $2$ respectively. The $0$'s in the last ket represents that there are no photons scattered at slit $1$, hence the electron is (initially) still in a superposition of two states (either at slit $1$ or slit $2$). $a$ and $u$ are parameters that initially are $1$, {\it i.e., } at $t=0$.

\noindent The various perturbed states are as follows.  In all these states, if a scattering event has occurred relative to slit $1$, the electron is now firmly in state $\ket{1}_e$. In fact, the state $\ket{2}_e \ket{n_0=1; n_1, n_2, ..., n_R}$ is not part of the possible Fock space of the system - physically, we cannot have a scattering event at slit $1$ while the electron is at slit $2$! We could, of course, expect noise, but that can be measured prior to the experiment and appropriately adjusted for. {\it {Hence, the only possible state of the system with the electron at slit $2$ is actually the second part of the state in Equation (6)}}.

\noindent The first of the perturbed states is where the initial photon is scattered, while the others are where subsequent photomultiplier events have occurred with the further production of other photons. These states are
\begin{eqnarray}
\ket{\psi}_1 \equiv  \ket{\psi}_{1,1;0...0} = \frac{\ket{1}_e}{\sqrt{2}} \ket{n_0=1; n_1=0, n_2 = 0, ..., n_R=0} \: \: \: \: \: \: \: \: \: \: \: \: \:  \: \: \: \: \: \: \: \: \: \: \: \: \: \:  \: \: \: \: \: \: \: \: \: \: \: \: \: \nonumber \\
\ket{\psi}_2^i \equiv \ket{\psi}_{1,0;0...N...0} = \frac{\ket{1}_e}{\sqrt{2}} \ket{n_0=0; n_1=0, n_2 = 0, .,n_i=N,.., n_R=0} \: \: \: \:  \: \: \: \: \: \: \: \: \: \: \: \: \: \nonumber \\
. \: \: \:  \: \: \: \: \: \: \: \: \: \: \: \: \: \: \: \: \: \: \: \: \: \: \: \: \: \: \: \: \: \: \: \: \: \: \: \: \: \: \: \: \: \: \: \: \: \: \: \: \: \: \: \nonumber \\
. \: \: \:  \: \: \: \: \: \: \: \: \: \: \: \: \: \: \: \: \: \: \: \: \: \: \: \: \: \: \: \: \: \: \: \: \: \: \: \: \: \: \: \: \: \: \: \: \: \: \: \: \: \: \:  \nonumber \\
. \: \: \:  \: \: \: \: \: \: \: \: \: \: \: \: \: \: \: \: \: \: \: \: \: \: \: \: \: \: \: \: \: \: \: \: \: \: \: \: \: \: \: \: \: \: \: \: \: \: \: \: \: \: \: \nonumber \\
\ket{\psi}_3^{ij} \equiv \ket{\psi}_{1,0;0,..., n_i=N-1,...n_j=M_1,...,0} \: \: \: \: \: \: \: \: \: \: \: \:  \: \: \: \: \: \: \: \: \: \: \: \: \: \: \:  \: \: \: \: \: \: \: \: \: \: \: \: \: \: \:  \: \: \: \: \: \: \: \: \: \: \: \: \: \: \:  \: \: \: \: \: \: \: \: \: \: \: \: \: \: \:  \: \: \: \: \: \: \: \: \: \: \: \: \:\nonumber \\
= \frac{\ket{1}_e}{\sqrt{2}} \ket{n_0=0; n_1=0, ..., n_i=N-1,...,n_j=M_1,...,n_R=0} \nonumber \\
\ket{\psi}_4^{ijk} \equiv \ket{\psi}_{1,0;0..., n_i=N-2,...n_j=M_1,...,n_k=M_1,...,0} \: \: \: \: \: \: \: \: \: \: \: \: \: \: \: \: \: \: \: \: \: \: \: \: \: \: \: \: \: \: \: \: \: \: \: \: \: \: \: \: \: \: \: \: \: \: \: \: \: \: \: \: \: \: \: \: \: \: \: \: \: \: \: \: \: \: \: \:  \: \nonumber \\
= \frac{\ket{1}_e}{\sqrt{2}} \ket{n_0=0; n_1=0, ..., n_i=N-2,...,n_j=M_1,...,n_k=M_1,...,n_R=0} \: \: \:  \: \: \: \: \: \: \: \: \: \: \: \:  \: \: \: \nonumber \\
...
\end{eqnarray}
\noindent there are of course many more possible end states and we can construct the series for the system's state $\ket{\psi}$ with more terms, as in Equation (8) below. We assume, in what follows that the states are labeled by $i$ and there are a macroscopic number $R$ of them.

\noindent Chosen this way, these states are themselves orthogonal to each other and since they are in the occupation number representation, can be enumerated and are complete. In general, we can write the state of the system at any time as
\begin{eqnarray}
\ket{\psi} =  e^{-i \omega_e t} \ket{\psi}_0 + b e^{-i (\omega_e  + \omega_p) t} \ket{\psi}_1 + \sum_{i = 1}^R c_i e^{-i (\omega_e + N \omega_i )t} \ket{\psi}_2^i  \nonumber \\
 + \sum_{ij; i\ne j} d_{ij} \ket{\psi}_3^{ij} e^{-i(\omega_e + (N-1) \omega_i+M_1 \omega_j) t} \nonumber \\
 + \sum_{ijk; i \ne j, j \ne k, k  \ne i} d_{ijk} \ket{\psi}_4^{ijk} e^{-i(\omega_e + (N-2) \omega_i+M_1 \omega_j+ M_1 \omega_k t ) t} + ...
\end{eqnarray}
 \noindent  where we have explicitly included the ``free'' time-dependence of the states in the exponentials multiplying every term. Note that the parameter $a$ introduced in Equation (6) appears in $\ket{\psi}_0$ itself.
 
In the above, $a$ is the amplitude (and $|a|^2$ the probability, by the Born rule) that the state is still {\it un-collapsed} and localized to slit $1$. Similarly, $b$ is the amplitude (and $|b|^2$ the probability) that the electron's state has collapsed into that localized to slit $1$ and one photon has been scattered into the idealized photomultiplier. Carrying on, $c_i$ is the amplitude (and $|c_i|^2$ the probability) that the electron's state has localized to slit $1$, while $N$ photons have then been generated in state $i$ by the initially scattered photon). We can carry this forward, {\it viz.}, $d_{ij}$ is the amplitude (and $|d_{ij}|^2$ the probability) that the electron's state has collapsed, localized to slit $1$, while $N-1$ photons are in state $i$ in the photomultiplier tube and $M_1$ photons are in state $j$ in the tube, all produced by the amplification process that was sparked by the first scattered photon. The process goes on, involving increasing numbers of photons in the photomultiplier tube in various other states.

As we noted before, there is still the chance that the electron actually is localized to slit $2$ and that is represented by the second term in $\ket{\psi}_0$.
 
In Equation (6), $a$ is initially $1$. {\bf {Clearly, for us to consider the state $\ket{\psi}$ to have collapsed with the electron localized to slit $1$, we will need $|a|\rightarrow 0$, while the sum $|b|^2+\sum_i |c_i|^2 + ... \rightarrow 1$. }} Indeed, we would want this to happen rather quickly. This central result is what we will try to obtain throughout the analysis in the paper.
 
We can now write down Schrödinger's equation for this general state, remembering that the initial state is described by $a=1, u=1, b=0, c_i=0 \: \forall i, d_{\vec n}=0 \: \forall {\vec n}$.
 \begin{eqnarray}
 i \frac{\partial \ket{\psi}}{\partial t} = {\cal H} \ket{\psi}
 \end{eqnarray}
 We successively pre-multiply the Schrödinger's equation as follows with the bra-vectors corresponding to the below kets,
 \begin{enumerate}
 \item $\ket{2}_e  \: \ket{n_0=0; n_1=0, n_2=0, ... , n_R = 0}   e^{- i \omega_e t}$ 
 \item $\sqrt{2} \ket{1}_e  \: \ket{n_0=0; n_1=0, n_2=0, ... , n_R = 0}   e^{- i \omega_e t}$ 
\item $\sqrt{2} \ket{1}_e \ket{n_0=1; n_1=0, n_2 = 0, ..., n_R=0} e^{- i \omega_e t - i \omega_p t}$
\item $\sqrt{2}\ket{1}_e \ket{n_0=0; n_1=0, n_2 = 0, .,n_i=N,.., n_R=0} e^{- i \omega_e t - i N \omega_i t}$
\item $\sqrt{2}\ket{1}_e \ket{0; n_1= 0, ..., n_i=N-1,..., n_j=M_1, ...,n_R=0} e^{- i \omega_e t - i (N-1) \omega_i t-i M_1 \omega_j t}$
\item $\sqrt{2}\ket{1}_e \ket{0; n_1= 0, ..., n_i=N-2,..., n_j=M_1, ...,n_k=M_1, ...,n_R=0} e^{- i \omega_e t - i (N-2) \omega_i t-i M_1 \omega_j t-i M_1 \omega_k t}$ 
\end{enumerate}
 We can continue this process for all the successive parameters.
 
 After some algebra, we are left with
 \begin{eqnarray}
 i \frac{\partial u}{\partial t}= 0  \rightarrow u(t)=1 \forall t  \: \: \: \: \: \: \: \: \: \: \: \: \: \: \: \: \: \: \: \: \: \: \: \: \: \: \: \: \: \: \: \: \: \: \: \: \: \: \: \: \: \: \: \: \: \: \: \: \: \: \: \: \: \: \: \: \:\: \: \: \: \: \: \: \: \: \: \: \: \: \: \: \: \: \: \: \: \: \: \: \: \: \: \: \: \: \: \: \: \: \: \: \: \nonumber \\
 i \frac{\partial a}{\partial t} = \Gamma_1 \: b \: \: \: \: \: \: \: \: \: \: \: \: \: \: \: \: \: \: \: \: \:\: \: \: \: \: \: \: \: \: \: \: \: \: \: \: \: \: \: \: \: \: \: \: \: \: \: \: \: \: \: \: \: \: \: \: \: \: \: \: \: \: \: \: \: \: \: \: \: \: \: \: \: \: \: \: \: \: \: \: \: \: \:\: \: \: \: \: \: \: \: \: \: \: \: \: \: \: \: \: \: \: \: \: \: \: \: \: \: \: \: \: \: \: \: \nonumber \\
  i \frac{\partial b}{\partial t} =  \Gamma_1 \: a  + \sum_{i=1}^R \Gamma_2^{(i)} \sqrt{N!} c_i  \: \: \: \: \: \: \: \: \: \: \: \: \: \: \: \: \: \: \: \: \: \: \: \: \: \: \: \: \: \: \: \: \: \: \: \: \: \: \: \:\: \: \: \: \:\: \: \: \: \: \: \: \: \: \: \: \: \: \: \: \: \: \: \: \: \: \: \: \: \: \: \: \: \: \: \: \: \: \: \: \: \: \nonumber \\
   i \frac{\partial c_i}{\partial t} = \Gamma_2^{(i)} \sqrt{N!} \:  b  +  \sum_{j=1; j \ne i}^{R} s \sqrt{N} \sqrt{M!} \: d_{ij} \: \: \: \: \: \: \: \: \: \: \: \: \: \: \: \: \: \: \: \: \: \: \: \:  \: \: \: \: \:  \: \: \: \: \: \: \: \: \: \: \: \: \: \: \: \: \: \: \: \: \: \: \: \: \: \: \: \: \nonumber \\
    i \frac{\partial d_{ij}}{\partial t} = s \sqrt{N} \sqrt{M_1!} \: c_i  + \sum_{k <>i,j} s \sqrt{N-1} \sqrt{M_1!} \: d_{ijk}\: \: \: \: \: \: \: \: \: \: \: \: \: \: \: \: \: \: \: \: \: \: \: \: \: \: \: \: \: \: \: \: \: \: \: \: \:  \nonumber \\
    ... \: \: \: \: \: \: \: \: \: \: \: \: \: \: \: \: \: \: \: \: \: \: \: \: \: \: \: \: \: \: \: \: \: \: \: \: \: \: \: \: \: \: \: \: \: \: \: \: \: \: \: \: \: \: \: \: \: \: \: \: \: \: \: \: \: \: \: \: \: \: \: \: \: \: \: \: \: \: \: \: \: \: \: \: \: \: \: \: \: \: \: \: \: \: \: \: \: \: \: \: \: \: \: \: \: \: \: \: \: \: \: \: \: \: \: \: \: 
 \end{eqnarray}
 where the ellipsis indicates that we can continue and write equations for more parameters in the state expansion.
 
\noindent In particular, generalizing, we could write the equations above (ignoring the simple solution for $u$) in a convenient matrix form as 
 \begin{eqnarray}
  i \frac{\partial a}{\partial t} = \Gamma_1 \: b \: \: \: \: \: \: \: \: \: \: \: \: \: \: \: \: \: \: \: \:\: \: \: \: \: \: \: \: \: \: \: \: \: \: \: \: \: \: \: \: \: \: \: \: \: \: \: \: \: \: \: \: \: \: \: \: \: \: \: \: \: \: \: \: \: \: \: \: \: \: \: \: \: \: \: \: \: \: \: \: \: \:\: \: \: \: \: \: \: \: \: \: \: \: \: \: \: \: \: \: \: \: \: \: \: \: \: \: \: \: \: \: \: \: \: \: \nonumber \\
 i \frac{\partial  b}{\partial t} =   \Gamma_1 \:  a  \:  + \sqrt{N!} \: { \bar \Gamma_2} .{\bar c} \: \: \: \: \: \: \: \: \: \: \: \: \: \: \: \: \: \: \: \: \: \: \: \: \: \: \: \: \: \: \: \: \: \: \: \: \: \: \: \: \: \: \: \: \: \: \: \: \: \: \: \: \: \: \: \: \: \: \: \: \: \: \: \: \: \: \: \: \: \: \: \: \: \: \: \: \: \: \: \: \: \: \: \: \: \: \: \: \: \: \: \: \nonumber \\
i \frac{\partial \bar c}{\partial t} =    {\bar \Gamma_2} \sqrt{N!} \:  b  \:  + s \sqrt{N}\: \sqrt{M_1!} \:   {\bar { \bar {\cal B}}} .{\bar d_2}  \: \: \: \: \: \: \: \: \: \: \: \: \: \: \: \: \: \: \: \: \: \: \: \: \: \: \: \: \: \: \: \: \: \: \: \: \: \: \:  \: \: \: \: \: \: \: \: \: \: \: \: \: \: \: \: \: \: \: \: \: \: \: \: \: \: \: \: \: \: \nonumber \\
i \frac{\partial \bar d_2}{\partial t} = s \sqrt{N} \: \sqrt{M_1!} \:{\bar {\bar {\cal B}}} . \: \bar c  + \sqrt{N-1} \sqrt{M_1!} \:{\bar {\bar {\cal E}}} . {\bar d_3} \: \: \: \: \: \: \: \: \: \: \: \: \: \: \: \: \: \: \: \: \: \: \: \: \: \: \: \: \: \: \: \: \: \: \: \: \: \: \: \: \: \: \: \: \: \: \: \: \: \nonumber \\
... \: \: \: \: \: \: \: \: \: \: \: \: \: \: \: \: \: \: \: \: \: \: \: \: \: \: \: \: \: \: \: \: \: \: \: \: \: \: \: \: \: \: \: \: \: \: \: \: \: \: \: \: \: \: \: \: \: \: \: \: \: \: \: \: \: \: \: \: \: \: \: \: \: \: \: \: \: \: \: \: \: \: \: \: \: \: \: \: \: \: \: \: \: \: \: \: \: \: \: \: \: \: \: \: \: \: \: \: \: \: \: \: \: \: \: \: \: 
\end{eqnarray}
 where
\begin{enumerate}
\item ${ \bar \Gamma_2}$ is a diagonal matrix, whose elements are the quantities $\Gamma_2^{(i)}$
\item ${\bar { \bar {\cal B}}}$ is a symmetric matrix, required by the unitarity conditions upon the time evolution. 
\item ${\bar {\bar {\cal E}}} $ is a symmetric matrix, again required by the unitarity conditions upon the time evolution. 
\end{enumerate}
We have used the notation $\bar c = (c_1, c_2, ..., c_R)$(the column vector)  and $\bar d_2 = (d_{ij \: \forall (i,j)})$, $\bar d_3 = (d_{ijk \: \forall (i,j,k)})$ (elements written as column vector whose elements are each distinct pair $(ij)$ and distinct triplet $(ijk)$ respectively) etc. Essentially, we write all the distinct states in a column-vector format. 

Note the structure of the equations; the coefficient of the term on the right side for every pair of parameters (say $(a,b)$) is the same.  The structure of the above is
 \begin{eqnarray}
i\frac{\partial}{\partial t} \left(\begin{array}{c} a \\
																						b \\
																						\bar c \\
																						\bar d_2 \\
																						\bar d_3 \\
																						... \end{array} \right)
= {\cal Q}_{total} \left(\begin{array}{c} a \\
																						b \\
																						\bar c \\
																						\bar d_2 \\
																						\bar d_3 \\
																						... \end{array} \right)  \nonumber
\end{eqnarray}
where ${\cal Q}_{total}$ is a time-independent hermitian matrix.
 
 \subsection{The strategy}
 
  We will pause for a bit to explain the strategy of using these equations to study our problem. We start things off, at $t=0$ with $a=1, u=1$ and all the other parameters set to $0$. This implies that the starting state is {\it uncollapsed} and the electron is in a superposition of two possible states, {\it viz.} localized to slit $1$ or slit $2$. Then we let the equation evolve in time, in completely unitary fashion. We will watch the sum of the squared moduli of $b, c_i, ...$, which we interpret, using the Born rule, as the total probability that the measuring apparatus has observed the electron, i.e., that the measuring apparatus has decohered the wave-function. This situation is consistent with a diagonal reduced density matrix \cite{Balian5}.
  
Now, since the measuring apparatus is involved in those ($b, c_i,...$) terms, it has also reduced it to the particular outcome that was produced from the precise sequence of microscopic interactions for that particular detection. In this sense, we can call this a ``collapse'' - the {\it yes/no} outcome has been settled in this one detection. 
  
Mathematically, the equations above are really just the equations to a collection of coupled harmonic oscillators and we are just watching one initially displaced oscillator transfer its energy to all the other oscillators.

\subsection{Study of the second order equations}
 
\noindent To simplify this set of equations, we operate on all of them with $i \frac{\partial}{\partial t}$ to yield the matrix equation (we can take the partial time derivative through ${\cal Q}_{total}$) below. We show only four elements in the column vector, of course, there are a huge number.
\begin{eqnarray}
- \frac{\partial^2}{\partial t^2} \left(\begin{array}{c} a \\
																						b \\
																						\bar c \\
																						\bar d_2 \end{array} \right) = {\cal Q}_{total}^2 \left(\begin{array}{c} a \\
																						b \\
																						\bar c \\
																						\bar d_2 \end{array} \right) \: \: \: \: \: \: \: \: \: \: \: \: \: \: \: \: \: \: \: \: \: \: \: \: \: \: \: \: \: \: \: \: \: \: \: \: \: \: \: \: \: \: \: \: \: \: \: \: \: \: \: \: \: \: \: \: \: \: \: \: \: 
 \nonumber \\
= \left(\begin{array}{cccc} {\bf \Gamma_1^2} & \: \: \:  0 &\: \: \:  \Gamma_1  {\bar \Gamma_2}  \sqrt{N!} &\: \: \:  0 \\
					     0 & \: \: \:  {\bf \Gamma_1^2+ N! \: {\bar \Gamma_2}.{\bar \Gamma_2} }& \: \: \: 0 & \: \: \:  s \sqrt{N M!} \: {\bar {\bar {\cal B}}} \\
					      \Gamma_1 {\bar \Gamma_2} \sqrt{N!} & \: \: \: 0 & \: \: \: {\bf N!  {\bar \Gamma_2}{\bar \Gamma_2}+ s^2 (N M!) {\bar {\bar {\cal B}}}.{\bar {\bar {\cal B}}}} &\: \: \: 0 \\
					     0 & \: \: \: s \sqrt{N M!} \: {\bar {\bar {\cal B}}} & \: \: \: 0 &\: \: \: {\bf s^2 (N M!) {\bar {\bar {\cal B}}}.{\bar {\bar {\cal B}}} }  \end{array}\right) \left(\begin{array}{c} a \\
																						b \\
																						\bar c \\
																						\bar d_2 \end{array} \right) 
\end{eqnarray}
where we haven't shown more rows and columns of the $ {\cal Q}_{total}^2$ matrix beyond $4$. There will of course be a huge number of rows and columns, corresponding to all possible sets of photons in several states.

In addition, the matrix ${\cal Q}_{total}^2$ is a symmetric, explicitly time-independent matrix. Indeed, due to the physical separation of the successive interactions at parts of the photomultiplier, the matrix is structurally one-skip-tridiagonal and symmetric. In what follows, we study the properties of a simplified version of this set of equations, consistent with the symmetric nature of ${\bar {\bar {\cal B}}}$ etc., with new variable definitions.  We refer to the matrix we studied above as $\cal G$, which is a $Q \times Q$ matrix. The underlying variables are renamed - $q_1$ corresponds to our $a$, while $q_2,...,q_Q$ to the other variables. Then,
\begin{eqnarray}
w = \left(\begin{array}{c} q_1 \\
																						q_2 \\
																						q_3 \\
																						q_4 \\
																						. \\
																						. \\
																						. \\
																						q_i \\
																						. \\
																						. \\
																						. \\
																						. \\
																						q_Q \end{array} \right)  \: \: \: \: \:  \: \: \: \: \: \rightarrow \: \: \: \: \: \: \: \: \: \:  - \frac{\partial^2}{\partial t^2} w
																						=  {\cal G}\:  w \nonumber
\end{eqnarray}

where the matrix $\cal G$ is
\begin{eqnarray}
{\cal G}= \left(\begin{array}{cccccccccc} a_1^2 & \: \: \:  0 &\: \: \:  a_1 a_2 & \: \: \: 0 \: \: \: &  \: \: \: 0 \: \: \: & ... & ... & ... &... &\: \: \: 0 \: \: \: \\
					     0 & \: \: \:  a_1^2+a_2^2 & \: \: \: 0 &  \: \: \: a_2 a_3 \: \: \: &\: \: \:  0 &  ... & ... & ... &... &0 \\
					     a_1 a_2 & \: \: \: 0 & \: \: \: a_2^2+a_3^2 & \: \: \: 0 \: \: \: &a_3 a_4 &   ... & ... &... &... &0 \\
					     0 & \: \: \: a_2 a_3 & \: \: \: 0 & \: \: \: a_3^2+a_4^2 \: \: \: &\: \: \: 0 &   ... & ... &... &... &0 \\
					     0 & \: \: \: 0 & \: \: \: a_3 a_4 & \: \: \: 0 \: \: \:   &\: \: \: a_4^2+a_5^2 &   ... & ... &... &... &0 \\
					     0 & \: \: \: 0 & \: \: \: 0 & \: \: \: a_4 a_5 \: \: \: &\: \: \: 0 & ... & ... &   ... &... &0 \\
					     0 & \: \: \: 0 & \: \: \: 0 & \: \: \: 0 \: \: \: &\: \: \: a_5 a_6 & ... & ... & ... & ... &0 \\
					     0 & \: \: \: 0 & \: \: \: 0 & \: \: \: 0 \: \: \: &\: \: \: ... & ... & ... & ... & ... &0 \\
					     0 & \: \: \: ... & \: \: \: ... & \: \: \: ... \: \: \: &\: \: \: ... & ... & ... & ... & ... &0 \\
					     0 & \: \: \: ... & \: \: \: ... & \: \: \: ... \: \: \: &\: \: \: ... & ... & ... & ... & ... &0 \\
					     0 & \: \: \: ... & \: \: \: ... & \: \: \: ... \: \: \: &\: \: \: ... & ... & ... & ... & ... &a_{Q-2} a_{Q-1} \\
					     0 & \: \: \: ... & \: \: \: ... & \: \: \: ... \: \: \: &\: \: \: ... & ... & ... & ... & ... &0 \\
					     0 & \: \: \: 0 & \: \: \: 0 & \: \: \: 0 \: \: \: &\: \: \: 0 & ... & ... & ... & ... &a_{Q-1}^2 
					      \end{array}\right)  \nonumber
\end{eqnarray}

The matrix ${\cal G}$ looks complicated, but its similarity to the matrix in Equation (12) is to be noted. We are studying a simplified form of Equation (12), as stated.

In the above matrix, we have, for instance, collapsed the set of components ${\underbar c}$ into one component $q_3$. Hence $a_2 \propto R$ and is, among other things, a macroscopic multiple of $\Gamma_2$ etc. The same goes for $a_5$ etc. We use $q_1$ to represent $a$ in the calculation, the starting amplitude for the electron to be {\it uncollapsed}, so its initial value (at $t=0$) is $1$. The other parameters $q_2, q_3, ...$  and their first derivatives w.r.t time are $0$ at $t=0$. Note that we need to specify the starting value of the coordinate as well as their rate of change since the equation is second-order in time.

In addition, $Q$ (the dimension of the problem) is all the possible kinds of states, in fact $Q \sim e^R$ since there are an exponential number of ways of partitioning cascaded photons amongst $R$ states.
 The one-skip-tridiagonal structure of the matrix has some interesting properties, which are very similar to the tri-diagonal structure seen in typical coupled oscillator problems. The eigenvalues of this matrix are positive semi-definite (see Appendix II). This means that the time evolution is purely unitary and no dissipation emerges from the mathematics of the problem. In addition, from an inspection of the equations we have obtained, the $a_i$'s increase by factors of roughly $R \sqrt{N}$ as $i$ increases.
 
For simplicity and for performing the numerical simulations, we consider $q_i$ chosen in a fairly simple manner to study the time evolution of a state that starts with $q_1|_{t=0}=1, \frac{\partial q_1}{\partial t}|_{t=0}=0$, i.e., the electron in a superposed state at both the slits. 
Accordingly, in Figures 2-5, we display the results of simulating the above equations for a simple form of the frequency matrix ${\cal G}$ chosen with $a_1=1, a_i=\sqrt{10} \: \: \forall \:  i \ne 1$. However, when we study the time-dependence of $q_1$, we will need to use a more realistic assumption for the $a_i$, namely, $a_1=a \: \: \& \: \:  a_i = a \:( R \:  \sqrt{N})^{i-1} \: \forall i>1$, as may be noted by looking at Equation (12). For clarity, $R$ is the macroscopic number of possible modes available at the first step (inside the photomultiplier) and $N$ is the relatively small (but greater than 1) number of photons produced at each stage of the photomultiplier cascade. 

This analysis is pretty robust. We can add small random terms to the elements of the matrix in a manner consistent with leaving it real, symmetric and of the form we are studying; interestingly this does not amend the results we discuss below. In addition, use may be made of Gershgorin's \cite{Gersh} theorem to understand why the eigenvalues are confined (with the numbers in our example) to the range $(0,..,40)$. For the purposes of graphing results, we have used $10, 100, 1000, 10000$ oscillators in the below examples of accompanying figures.
 \begin{eqnarray}
 {\cal G} 
= \left(\begin{array}{cccccccccc} 1 & \: \: \:  0 &\: \: \:  \sqrt{10} & \: \: \: 0 \: \: \: &  \: \: \: 0 \: \: \: & ... & ... & ... &... &\: \: \: 0 \: \: \: \\
					     0 & \: \: \:  11 & \: \: \: 0 &  \: \: \: 10 \: \: \: &\: \: \:  0 &  ... & ... & ... &... &0 \\
					     \sqrt{10} & \: \: \: 0 & \: \: \: 20 & \: \: \: 0 \: \: \: &10 &   ... & ... &... &... &0 \\
					     0 & \: \: \: 10 & \: \: \: 0 & \: \: \: 20 \: \: \: &\: \: \: 0 &   ... & ... &... &... &0 \\
					     0 & \: \: \: 0 & \: \: \: 10 & \: \: \: 0 \: \: \:   &\: \: \: 20 &   ... & ... &... &... &0 \\
					     0 & \: \: \: 0 & \: \: \: 0 & \: \: \: 10 \: \: \: &\: \: \: 0 & ... & ... &   ... &... &0 \\
					     0 & \: \: \: 0 & \: \: \: 0 & \: \: \: 0 \: \: \: &\: \: \: 10 & ... & ... & ... & ... &0 \\
					     0 & \: \: \: 0 & \: \: \: 0 & \: \: \: 0 \: \: \: &\: \: \: ... & ... & ... & ... & ... &0 \\
					     0 & \: \: \: ... & \: \: \: ... & \: \: \: ... \: \: \: &\: \: \: ... & ... & ... & ... & ... &0 \\
					     0 & \: \: \: ... & \: \: \: ... & \: \: \: ... \: \: \: &\: \: \: ... & ... & ... & ... & ... &0 \\
					     0 & \: \: \: ... & \: \: \: ... & \: \: \: ... \: \: \: &\: \: \: ... & ... & ... & ... & ... &10 \\
					     0 & \: \: \: ... & \: \: \: ... & \: \: \: ... \: \: \: &\: \: \: ... & ... & ... & ... & ... &0 \\
					     0 & \: \: \: 0 & \: \: \: 0 & \: \: \: 0 \: \: \: &\: \: \: 0 & ... & ... & 10 & 0 &10 
					      \end{array}\right) 
 \end{eqnarray}
 
 \begin{figure}[h!]
\caption{10 oscillators, 10,000 time steps of 0.01 seconds each: Magnitude of $q_1$}
\centering
\includegraphics[scale=.7]{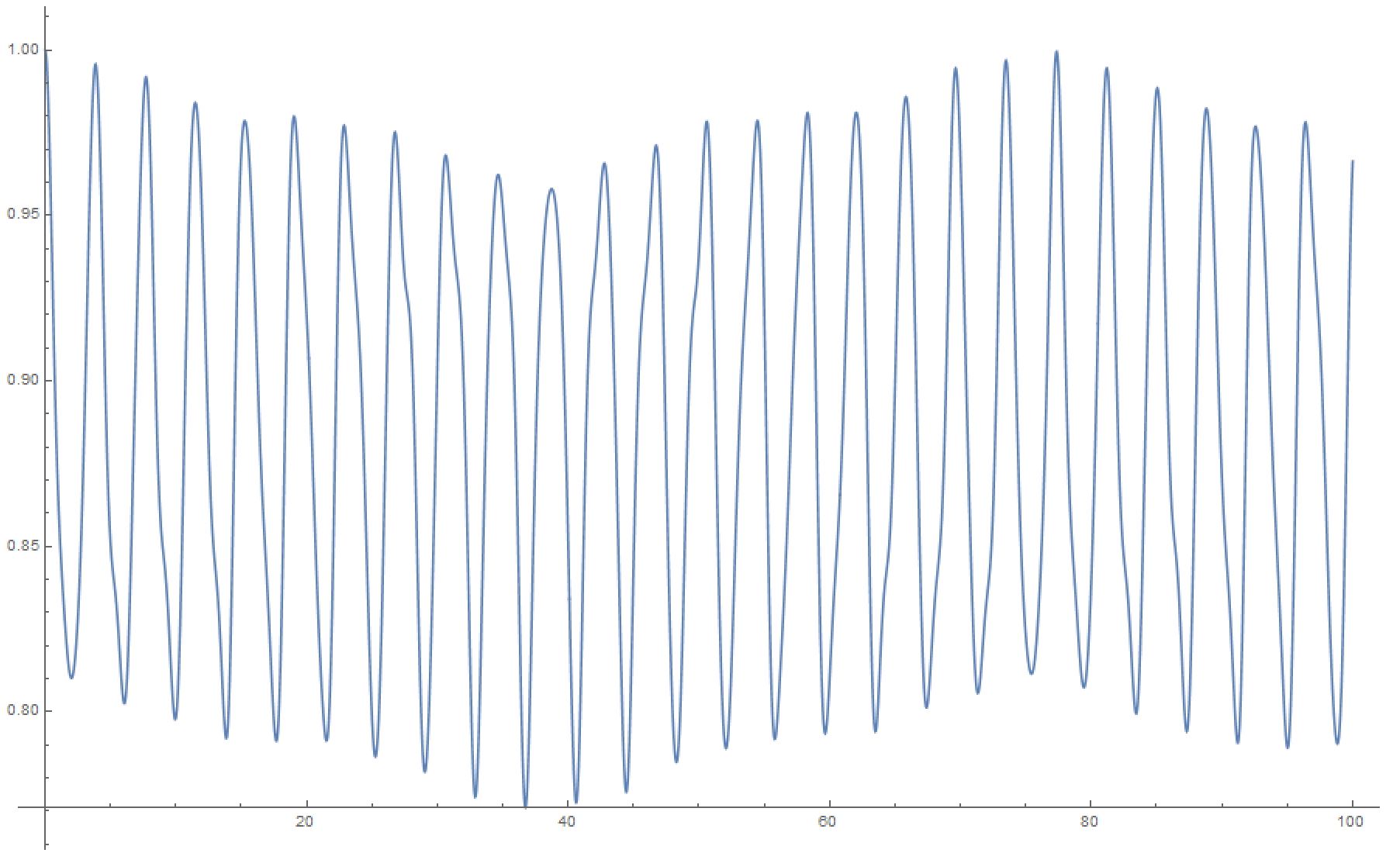}
\end{figure}

 \begin{figure}[h!]
\caption{100 oscillators, 10,000 time steps of 0.01 seconds each: Magnitude of $q_1$}
\centering
\includegraphics[scale=.7]{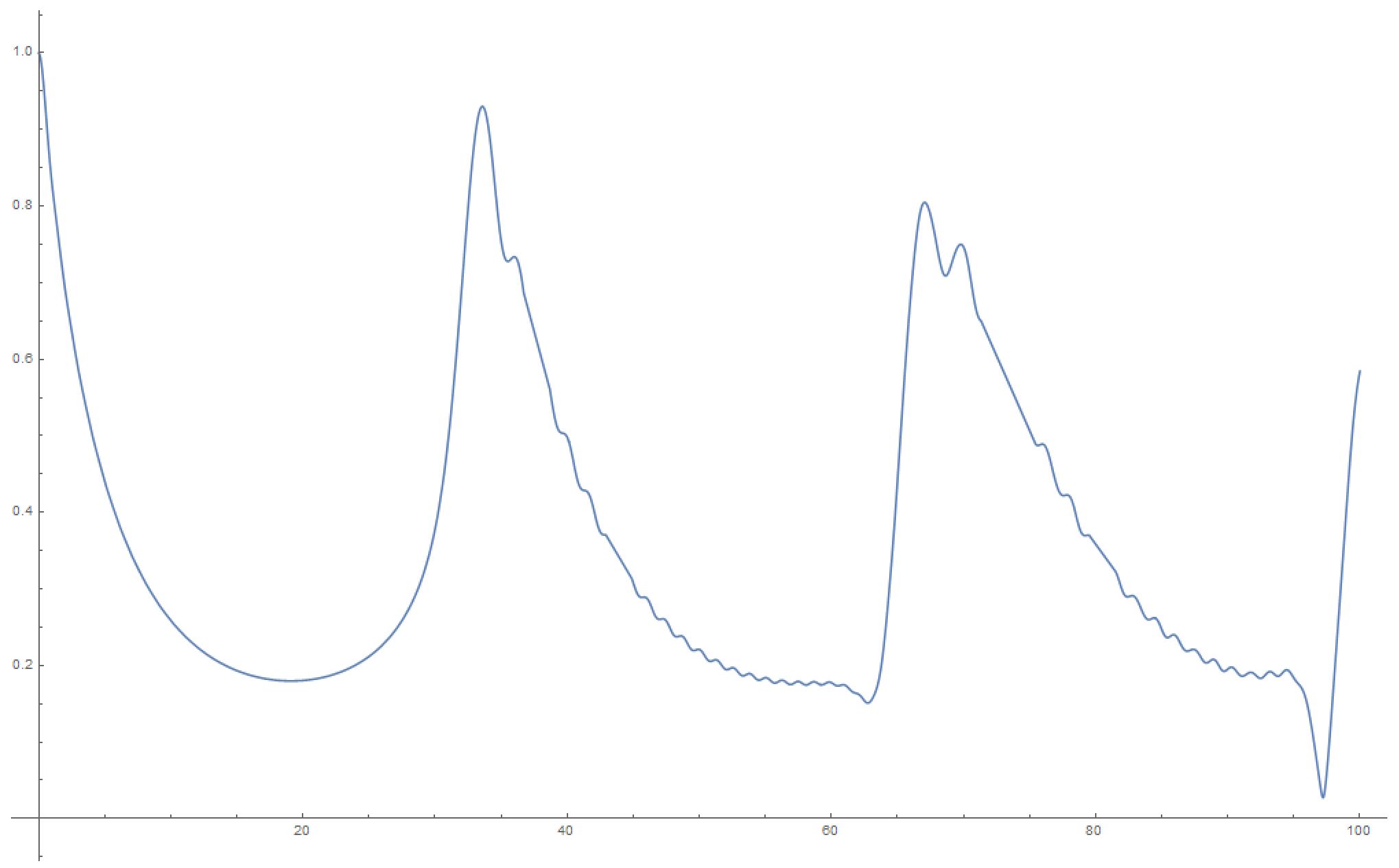}
\end{figure}
  
 \begin{figure}[h!]
\caption{1000 oscillators, 10,000 time steps of 0.01 seconds each: Magnitude of $q_1$}
\centering
\includegraphics[scale=.7]{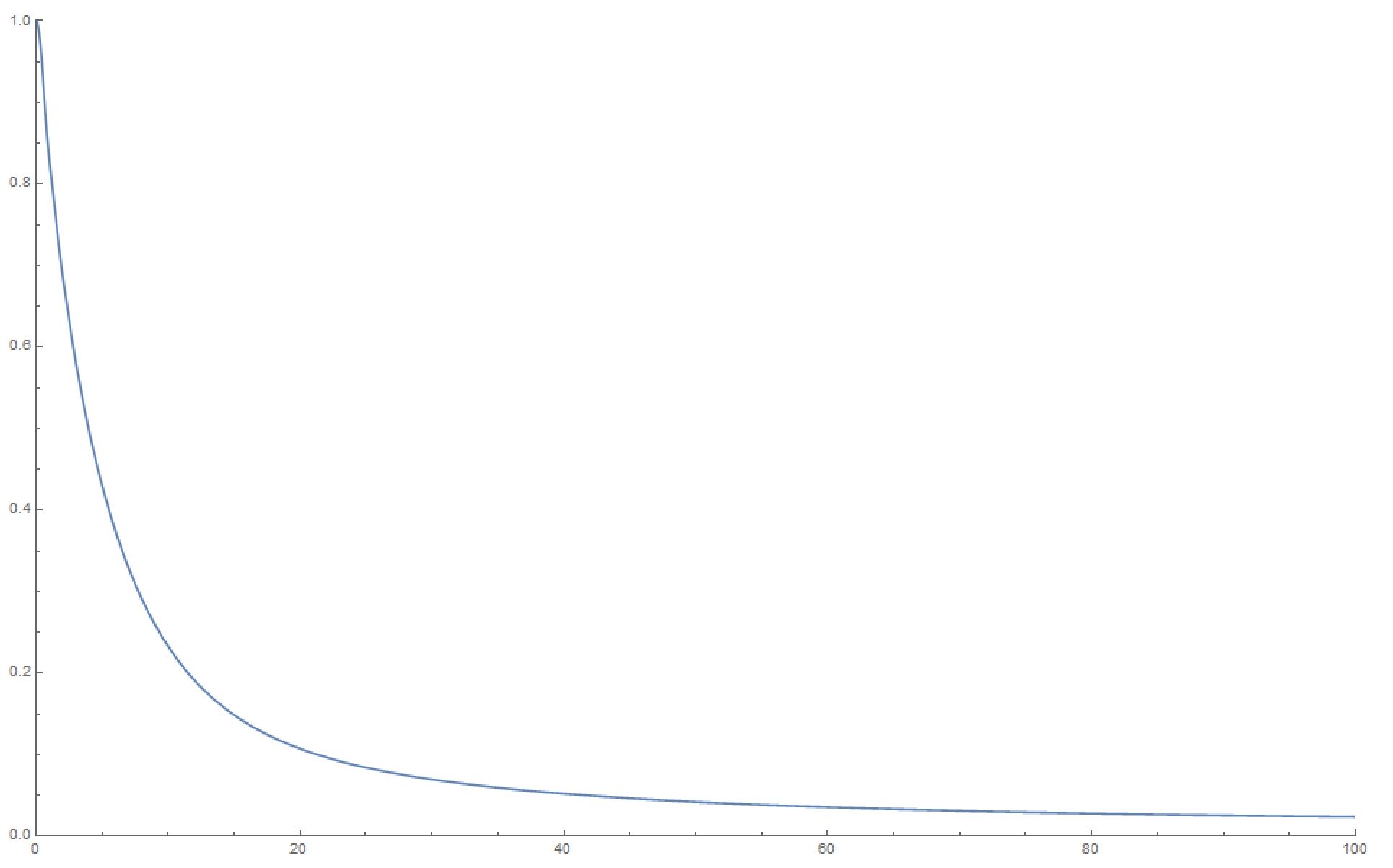}
\end{figure}
 
 \begin{figure}[h!]
\caption{10000 oscillators, 10,000 time steps of 0.01 seconds each: Magnitude of $q_1$}
\centering
\includegraphics[scale=.7]{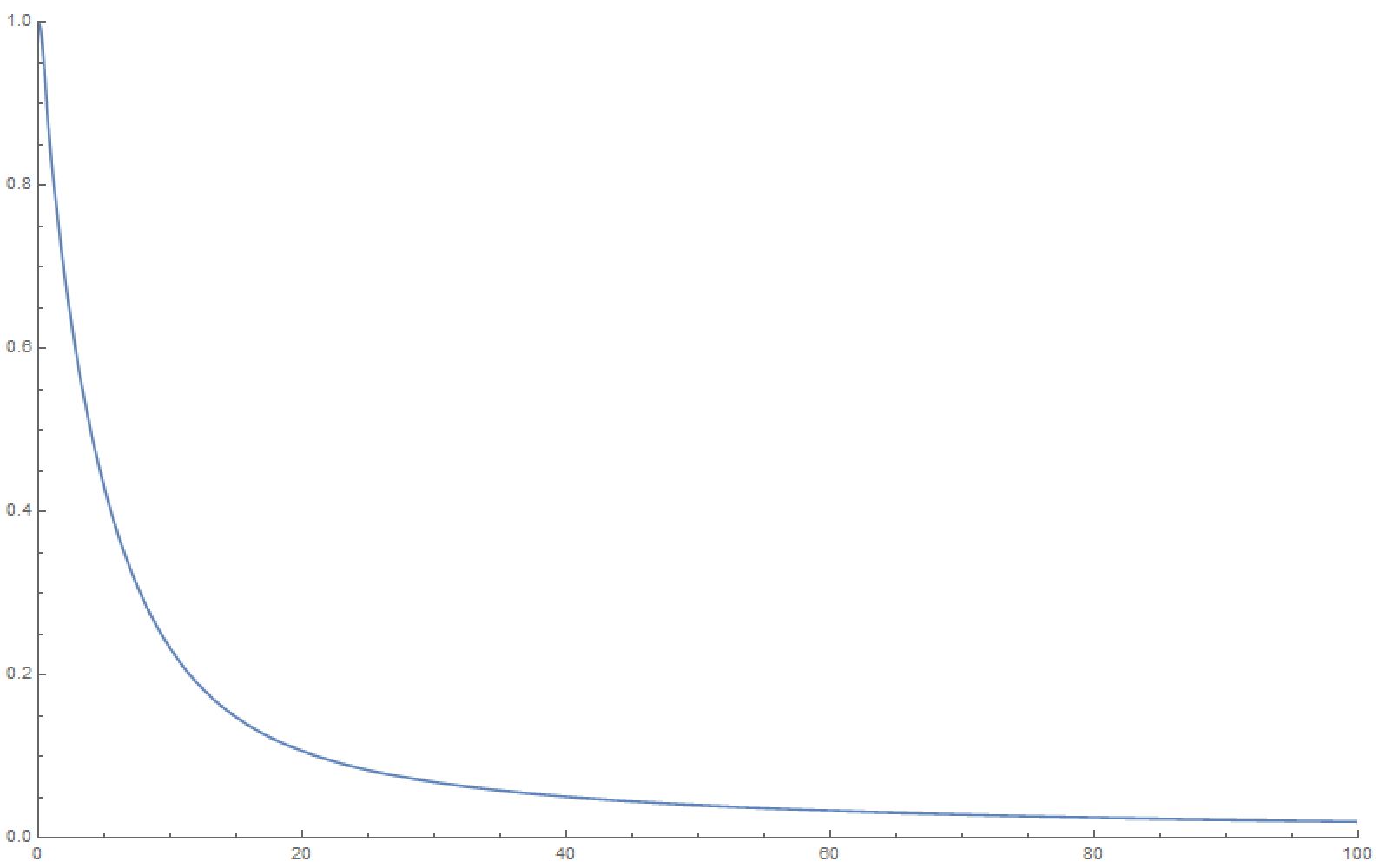}
\end{figure}
 
 To explain what's happening in the problem, the starting $q_1=1$ weight in the collection of coupled harmonic oscillators is distributed to the other oscillators due to the coupling between them. For a small enough set of oscillators, the behaviour is itself periodic and recurs. That is to say, the wave-function can {\it un-collapse} in a small enough system. In a large system, the time required to revisit the initial configuration can be very large, due to the macroscopic number of participating ``oscillators''. The system therefore stays {\it collapsed}.
 
In the next section, we will show that the behavior of $q_1$ for a large enough system is exponential, hence producing exponential decay for long time. However, for short times, the decay is quadratic in time (Fermi's golden rule). These results are simply structural, result from an analysis of the density of eigenvalues of the large $\cal G$ matrix (for the long-time limit), as well as the largest eigenvalue (for the short-time limit). The key result is Equation (27).

 \section{Obtaining decaying behavior from a collection of harmonic oscillators}
 
 To understand why we see behavior exhibited in Fig. 4 and Fig. 5, we analyze the problem as follows: the relevant eigenvalues and eigenvectors are those of the matrix $\cal G$. The normalized eigenvectors ($Q$ of them) of $\cal G$ are written as $\hat e_1, \hat e_2, ..., \hat e_Q$. We define the following matrices, in the original state vectors of the system as
 \begin{eqnarray}
 \hat {\cal V}^T = \left(\begin{array}{cccccc} ( . & . & . & \hat e_1 & . & . )\\
 ( . & . & . & \hat e_2 & . & . ) \\
 ( . & . & . & ... & . & . ) \\
  ( . & . & . & ... & . & . ) \\
 ( . & . & . & \hat e_Q & . & . )
 \end{array}\right) \nonumber \\
  \hat {\cal V} = \left(\begin{array}{cccccc} \Bigg( \hat e_1 \Bigg) & \Bigg( \hat e_2 \Bigg) & ... & . & . & \Bigg( \hat e_Q \Bigg)
 \end{array}\right) \nonumber \\
  \hat {\cal V^T}  \hat {\cal V} = {\cal I} \rightarrow  \hat {\cal V}^T =  \hat {\cal V}^{-1}
 \end{eqnarray}
 Also, define the weight vector $w$ for the eigenstates to produce the initial system state, as well as the initial system state ${\cal B}(t=0)$ as
 \begin{eqnarray}
 w = \left(\begin{array}{c} w_1\\
 w_2\\
.\\
 .\\
 w_Q
 \end{array}\right) \: \: \: \: \: 
  {\cal B}(t=0) = \left(\begin{array}{c} 1\\
 0\\
.\\
 .\\
 0 \end{array}\right)
 \end{eqnarray}
 We note that since $\cal B$ represents the initial state of the system, we must have
 \begin{eqnarray}
 {\cal V} . w = {\cal B}(t=0)
 \end{eqnarray}
 This yields, formally, the solution to the initial weights as $w = {\cal V}^T . {\cal B}(t=0)$. This implies, upon inspection, that
 \begin{eqnarray}
 w_i = {\hat e}_i^{(1)}
 \end{eqnarray}
 where we have used the notation ${\hat e}_i^{(1)}$ for the first element of eigenvector ${\hat e}_i$.
 In this notation, we have, for the time-evolution of $q_1(t)$, the amplitude of the initial state, 
 \begin{eqnarray}
 q_1(t) = \sum_{i=1}^Q {\hat e}_i^{(1)} {\hat e}_i^{(1)} \cos {\Omega_i t} = \sum_{i=1}^Q w_i^2  \cos {\Omega_i t}
 \end{eqnarray}
 This is obtained from the general solution to the state of the system ${\cal B}(t)$, which is,
 \begin{eqnarray}
 {\cal B}(t) = e^{i\sqrt{{\cal G}} t } {\cal V} w = e^{i\sqrt{{\cal G}} t } {\cal B}(t=0)
 \end{eqnarray}
 and applying the initial condition $q_1(t=0)=1$. Remember, we need to compute the future time dependence of $q_1(t)$.
 
To achieve this (see Equation (27)) let's now study the dependence of the weights $w_i$ upon the eigenvalues $\Omega_i^2$ in this simplified model: as we will discover, this relationship is what we need in order to demonstrate the results. In particular, we want the large $\Omega_i$ dependence of $w_i^2$. 

We begin with the eigenvalue equation for $\cal G$, i.e., ${\cal G} {\hat e}_i = \Omega_i^2 {\hat e}_i $. Writing down each term and noting that $w_i={\hat e}_i^{(1)}$, the first component of the normalized eigenvector, we get
\begin{eqnarray}
 a_1^2  {\hat e}_i^{(1)} + a_1 a_2 {\hat e}_i^{(3)} = \Omega_i^2 {\hat e}_i^{(1)} \nonumber \\
 a_1 a_2   {\hat e}_i^{(1)} + (a_2^2+a_3^2) {\hat e}_i^{(3)} + a_3 a_4 {\hat e}_i^{(5)}= \Omega_i^2 {\hat e}_i^{(3)} \nonumber \\
 ...
 \end{eqnarray}
 which can be solved generally as 
 \begin{eqnarray}
 {\hat e}_i^{(1)} = {\hat e}_i^{(2n+1)} \frac{a_1 a_2...a_{2n}}{\Omega_i^{2n} - \Omega_i^{2n-2} (a_1^2+a_2^2+...+a_{2n-1}^2)+...+ (-1)^n a_1^2 a_3^2...a_{2n-1}^2}
 \end{eqnarray}
 
 Note that we can write down a similar condition for ${\hat e}_i^{(2)}$. There are two potential eigenvectors that solve the problem for one $\Omega_i$ - one series with ${\hat e}_i^{(1)} =0$ (and all subsequent odd-index eigenvector components equal to 0) and the other with ${\hat e}_i^{(2)}=0$ (and so for the subsequent even-index weights). The solution with ${\hat e}_i^{(1)} =0$ has zero weight and so doesn't contribute to the value of $q_1(t)$ at {\it any} time $t$ and can be ignored.  We therefore need to only keep the odd-numbered components of the eigenvectors and the only eigenvectors that matter are the ones where the odd-numbered weights are non-zero. 
 
 In what follows, we use the more general form, i.e., $a_1=a, \: a_i = a\: R N^{\frac{i-1}{2}} \: \forall i>1$.
 For the largest eigenvalues, the formula for the eigenvector components can be approximated by 
 \begin{eqnarray}
 {\hat e}_i^{(2n+1)} \approx {\hat e}_i^{(1)}  \frac{\Omega_i^{2n}}{a_1 a_2...a_{2n}}  
 \end{eqnarray}
 Hence, the normalization condition for the eigenvector ${\hat e}_i$ is
 \begin{eqnarray}
 ({\hat e}_i^{(1)})^2 \bigg( 1 + (\frac{\Omega_i^2}{a_1 a_2})^2 + (\frac{\Omega_i^4}{a_1 a_2 a_3 a_4})^2 + ...\bigg) = 1
 \end{eqnarray}
 
\noindent Hence, approximating for the largest eigenvalues as $Q\rightarrow \infty$,
 \begin{eqnarray}
 ({\hat e}_i^{(1)})^2  \equiv w_i^2 \approx e^{- \frac{\Omega_i^4}{a_1^2 a_2^2}} = e^{- \frac{\Omega_i^4}{R^2 \: N a^4}} 
 \end{eqnarray} 
 
\noindent For small eigenvalues, the opposite limit can be applied, namely
 \begin{eqnarray}
{\hat e}_i^{(2n+1)} \approx  {\hat e}_i^{(1)} \frac{(-1)^{n-1} \Omega_i^2 (\sum_m^{`} A_m)+ (-1)^n a_1^2 a_3^2...a_{2n-1}^2}{a_1 a_2...a_{2n}}
 \end{eqnarray}
 where $A_m$ has  $(n-1)$ multiples of coefficients. Specifically, for the first four components, we note the pattern
 \begin{eqnarray}
 {\hat e}_i^{3} = {\hat e}_i^{1} \: \frac{\Omega_i^2 - a_1^2}{a_1 a_2} = {\hat e}_i^{1} \: \frac{\Omega_i^2 - a^2}{a^2 R \sqrt{N}}\: \: \: \: \: \: \: \: \: \: \: \: \: \: \: \: \: \: \: \: \: \: \: \: \: \: \: \: \: \:\: \: \: \: \: \: \: \: \: \: \: \: \: \: \: \: \: \nonumber \\
 {\hat e}_i^{5} = - {\hat e}_i^{1} \: \frac{\Omega_i^2 - \frac{a_1^2a_3^2}{a_1^2+a_2^2+a_3^2}}{\frac{a_1 a_2 a_3 a_4}{a_1^2+a_2^2+a_3^2}} \approx - {\hat e}_i^{1} \: \frac{\Omega_i^2 - \frac{a^2}{3}}{\frac{R^2 N \: a^2}{3}} \: \: \: \: \: \: \: \: \: \: \: \: \: \: \: \: \: \: \: \: \: \: \: \: \: \: \: \: \: \:\: \: \: \nonumber \\
 {\hat e}_i^{7} = {\hat e}_i^{1} \: \frac{\Omega_i^2 - \frac{a_1^2a_3^2 a_5^2}{a_1^2a_3^2+a_1^2 a_4^2+a_1^2 a_5^2+a_2^2a_4^2+a_2^2a_5^2+a_3^2a_5^2}}{\frac{a_1 a_2 a_3 a_4 a_5 a_6}{a_1^2a_3^2+a_1^2 a_4^2+a_1^2 a_5^2+a_2^2a_4^2+a_2^2a_5^2+a_3^2a_5^2}} \approx {\hat e}_i^{1} \: \frac{\Omega_i^2 - \frac{a^2}{6}}{ \frac{R^3 N^{3/2}a^2}{6}} \: \: \: \:\nonumber \\
  {\hat e}_i^{9} = - {\hat e}_i^{1} \: \frac{\Omega_i^2 - \frac{a_1^2a_3^2 a_5^2 a_7^2}{a_1^2a_3^2(a_5^2+a_6^2+a_7^2) +(a_1^2 +a_2^2+a_3^2) a_5^2 a_7^2+a_1^2 a_4^2 (a_6^2+a_7^2) + a_2^2 a_4*2 (a_6^2+a_7^2)}}{\frac{a_1 a_2 a_3 a_4 a_5 a_6 a_7 a_8}{a_1^2a_3^2(a_5^2+a_6^2+a_7^2) +(a_1^2 +a_2^2+a_3^2) a_5^2 a_7^2+a_1^2 a_4^2 (a_6^2+a_7^2) + a_2^2 a_4*2 (a_6^2+a_7^2)}} \approx - {\hat e}_i^{1} \: \frac{\Omega_i^2 - \frac{a^2}{16}}{\frac{R^4 N^2 a^2}{16}}
 \end{eqnarray}
 Note the $R\sqrt{N}$ factors that grow in the denominator and we have replaced terms in the denominator by the largest (in each case) in our approximations for ${\hat e}_i^{5}, {\hat e}_i^{7}, ...$.
 
Again, using the normalization condition for the eigenvector ${\hat e}_i$, the small eigenvalue expansion leads to 
 \begin{eqnarray}
 ({\hat e}_i^{(1)})^2 \bigg( 1 + \frac{1}{R^2N} \big(\frac{\Omega_i^2-a^2}{a^2 } \big)^2 + \frac{1}{R^4N^2} \big(\frac{\Omega_i^2-\frac{a^2}{3}}{\frac{a^2}{3}} \big)^2 + \frac{1}{R^6N^3} \big(\frac{\Omega_i^2-\frac{a^2}{6}}{\frac{a^2}{6}} \big)^2 + \frac{1}{R^8N^4} \big(\frac{\Omega_i^2-\frac{a^2}{16}}{\frac{ a^2}{16}} \big)^2 +..\bigg)=1 \nonumber \\
 \rightarrow  ({\hat e}_i^{(1)})^2  \equiv w_i^2  \approx \frac{1}{\bigg(1 +  \frac{1}{R^2N} (\frac{ \Omega_i^2}{a^2} -1)^2\bigg) } \nonumber
 \end{eqnarray}
 where we have kept only the first term in the set of $Q$ terms in the denominator.

To make some numerical estimates, we plot the squared weights and eigenvalues for the various finite-sized systems for our simple choice of the $\cal G$ matrix in Equation (13), in Figures 6-13.

 \begin{figure}[h!]
\caption{Squared Weights: 10 oscillators, 10,000 time steps of 0.01 seconds each}
\centering
\includegraphics[scale=.7]{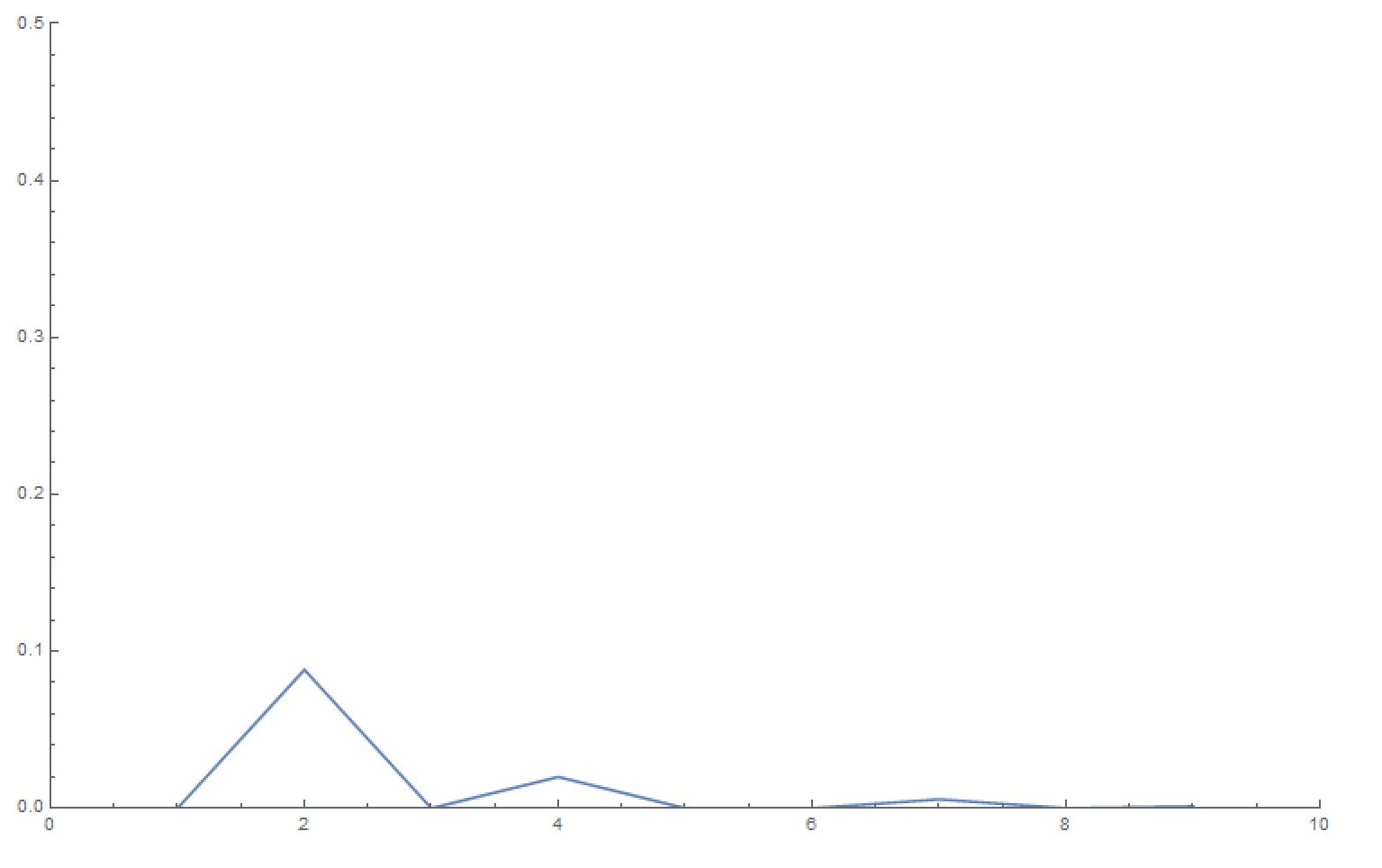}
\end{figure}

 \begin{figure}[h!]
\caption{Eigenvalues: 10 oscillators, 10,000 time steps of 0.01 seconds each}
\centering
\includegraphics[scale=.7]{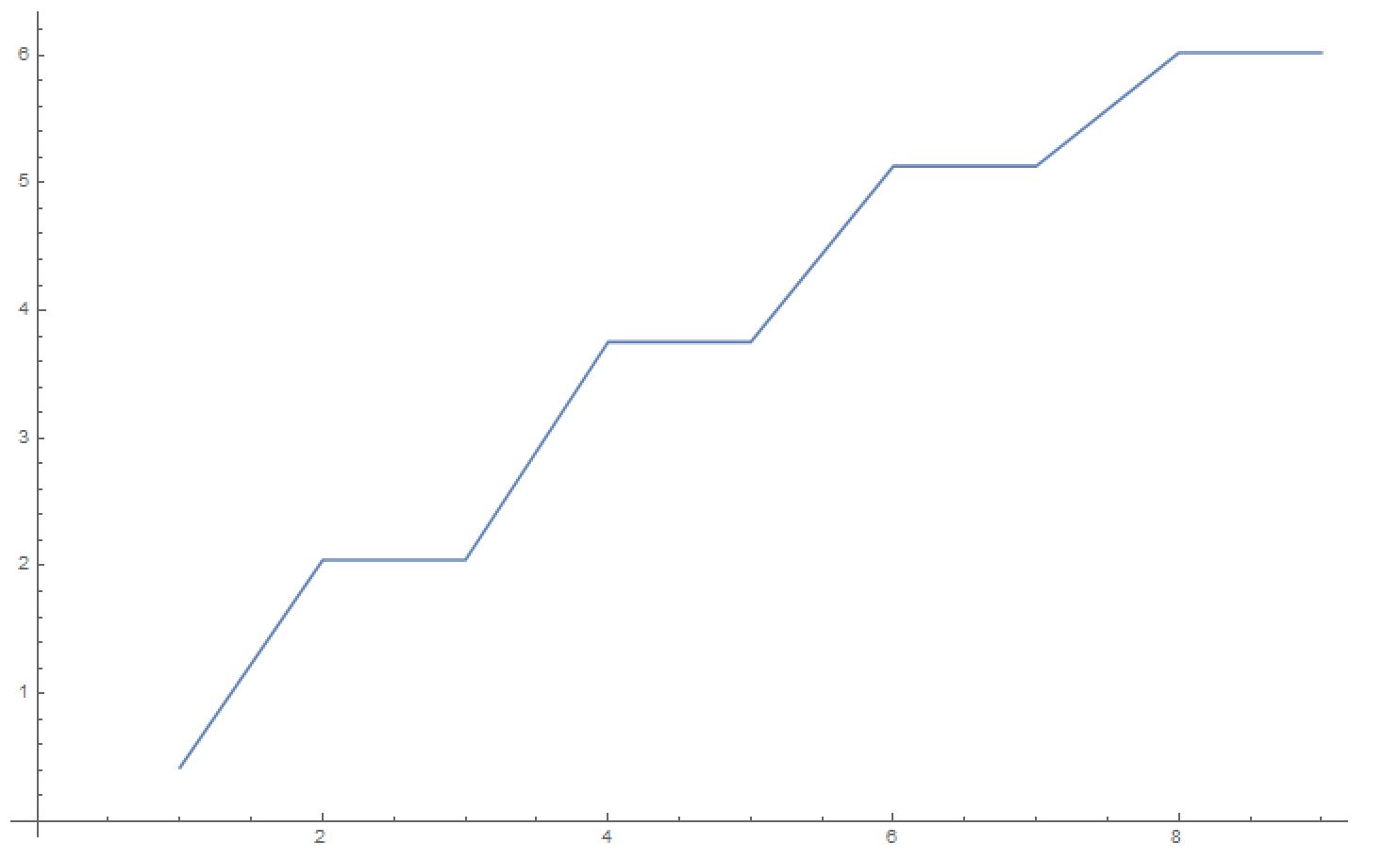}
\end{figure}

 \begin{figure}[h!]
\caption{Squared Weights: 100 oscillators, 10,000 time steps of 0.01 seconds each}
\centering
\includegraphics[scale=.7]{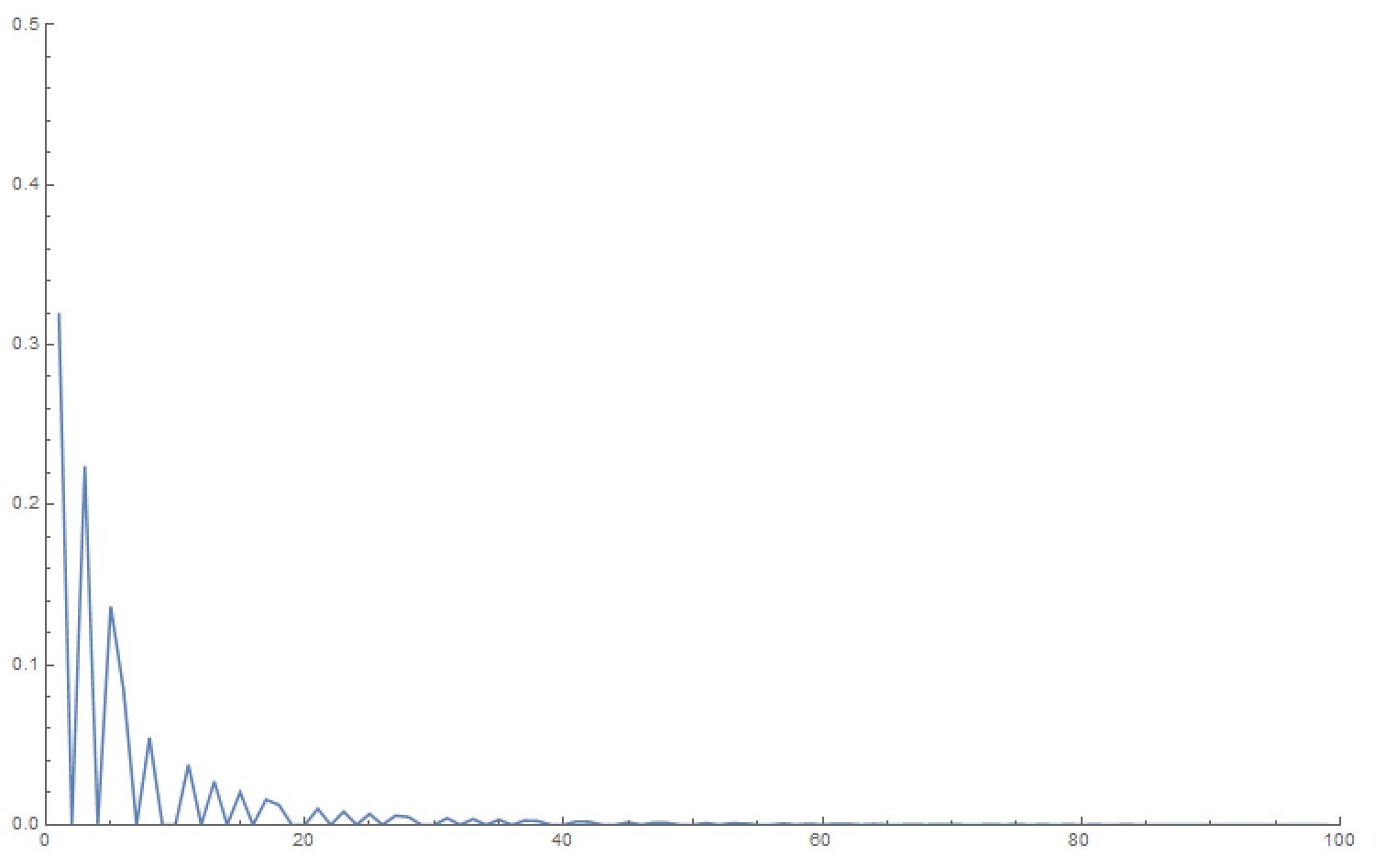}
\end{figure}

 \begin{figure}[h!]
\caption{Eigenvalues: 100 oscillators, 10,000 time steps of 0.01 seconds each}
\centering
\includegraphics[scale=.7]{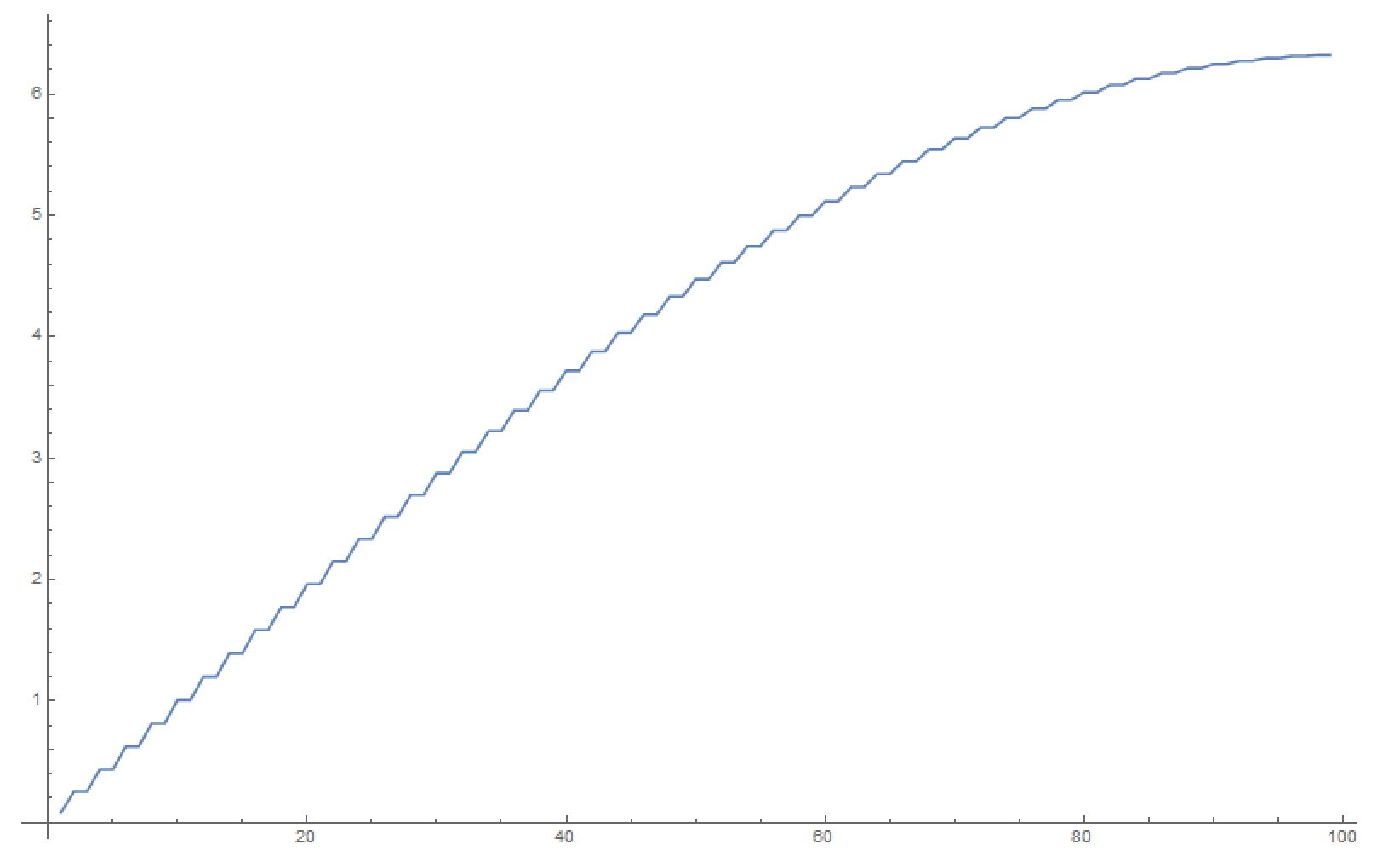}
\end{figure}

 \begin{figure}[h!]
\caption{Squared Weights:1000 oscillators, 10,000 time steps of 0.01 seconds each}
\centering
\includegraphics[scale=.7]{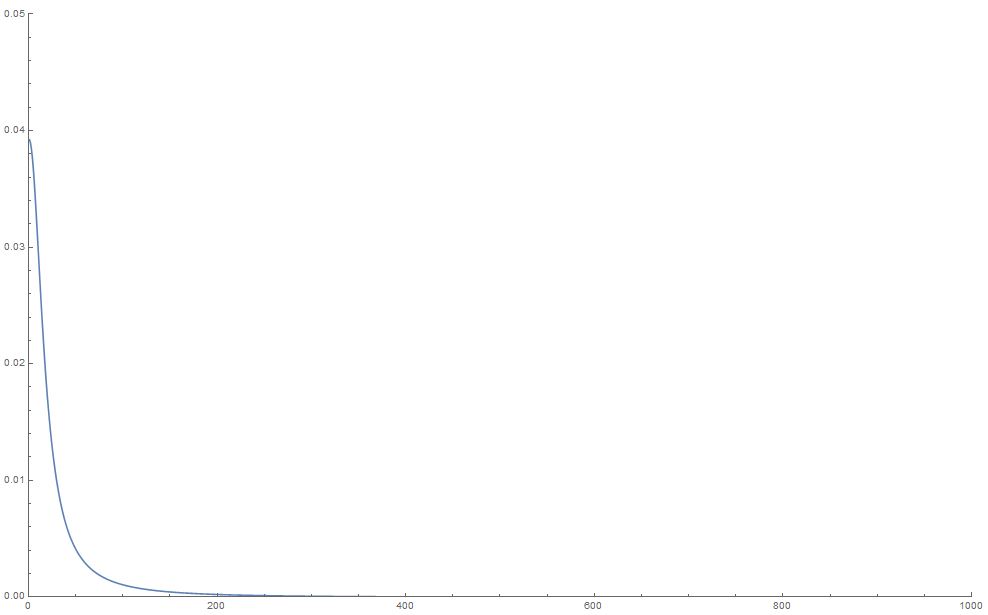}
\end{figure}

 \begin{figure}[h!]
\caption{Eigenvalues: 1000 oscillators, 10,000 time steps of 0.01 seconds each}
\centering
\includegraphics[scale=.7]{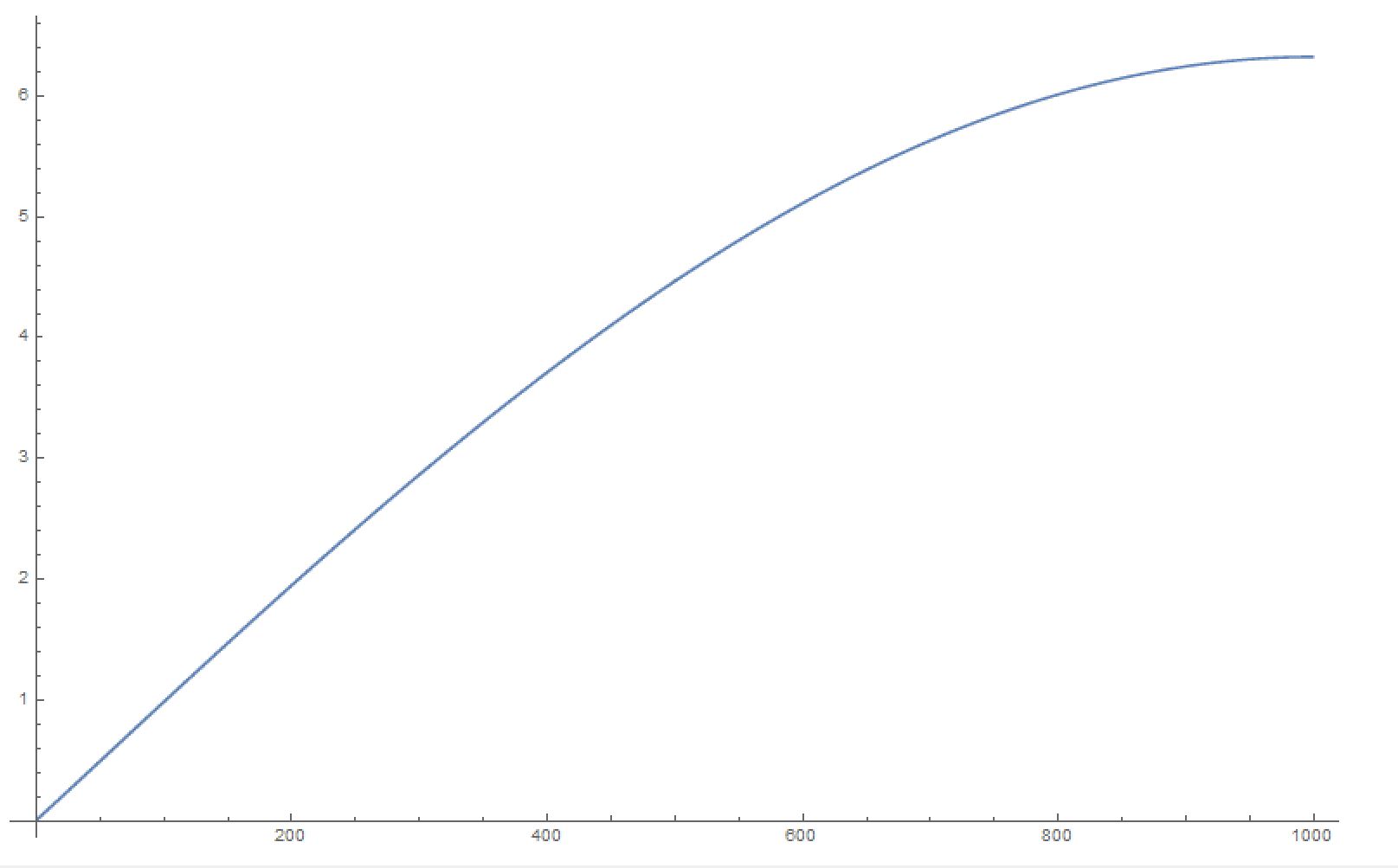}
\end{figure}

 \begin{figure}[ht!]
\caption{Squared Weights: 10000 oscillators, 10,000 time steps of 0.01 seconds each}
\centering
\includegraphics[scale=.7]{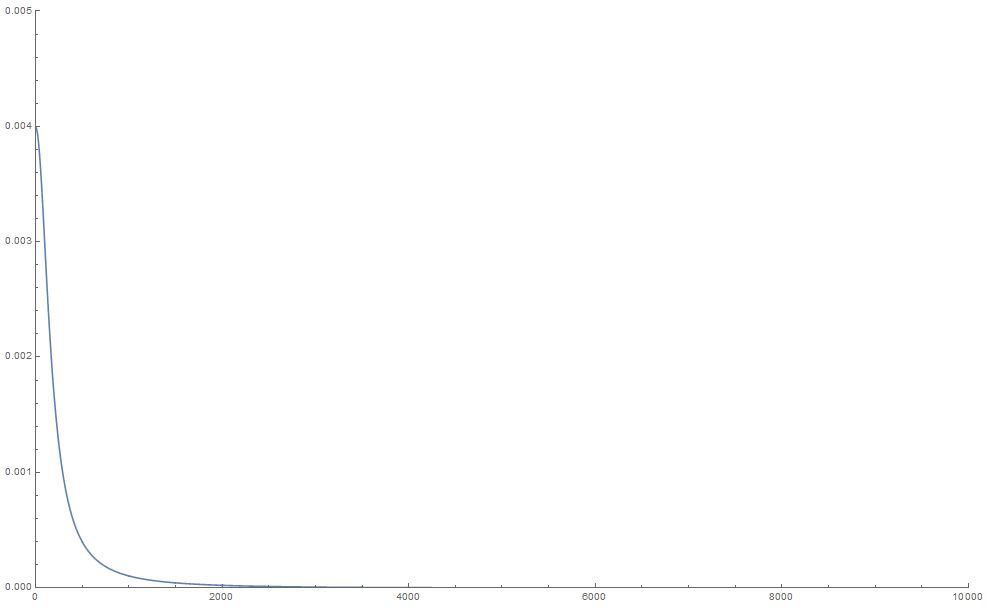}
\end{figure}

 \begin{figure}[ht!]
\caption{Eigenvalues: 10000 oscillators, 10,000 time steps of 0.01 seconds each}
\centering
\includegraphics[scale=.7]{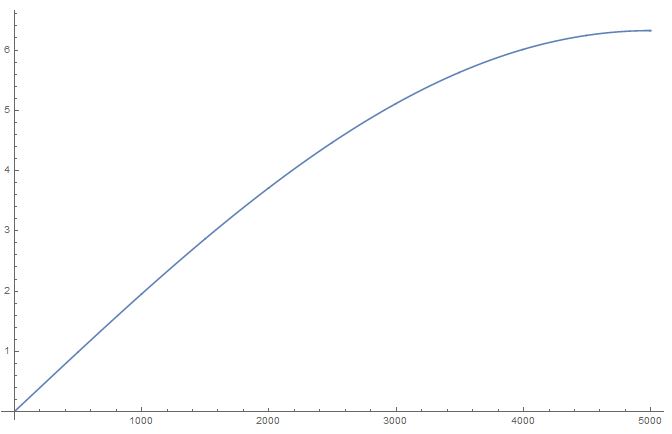}
\end{figure}

\noindent The time-dependence of $q_1(t)$ is written finally as with the functional forms we have derived in the previous section. 

We write, in the limit of an infinite number of oscillators and using the chain rule to write the sum over each of the eigenvalues as an integral over the actual eigenvalues,
 \begin{eqnarray}
 q_1(t) = \sum_{i=1}^Q w_i^2  \cos {\Omega_i t} \approx  \int_0^{\infty} d \Omega \: {\cal D}(\Omega) \: \:  w(\Omega)^2 \cos {\Omega t} 
 \end{eqnarray}
 From an inspection of the density of eigenvalues, we note that it is approximately constant (the slope of the $i$-vs-$\Omega_i$ curves, see Fig. 11), hence we can use ${\cal D}(\Omega) = {\cal C}$.

For short times, we use the large eigenvalue limit of the weight/eigenvalue dependence, i.e.,
 \begin{eqnarray}
 q_1(t)|_{ST} \approx {\cal C} \int_0^{\infty} d \Omega \: \:  e^{- \frac{\Omega^4}{R^2 N a^4} } \cos {\Omega t} 
 \end{eqnarray}
 where ${\cal C}$ must be set such that we recover $ q_1(t=0)|_{ST} =1$ from the integral approximation. The time dependence is therefore, essentially, the Fourier transform of the squared weights as a function of the eigenvalues.
 To estimate the behavior of this integral, we note the two limits \cite{Boyd}, namely,
 \begin{eqnarray}
\int_0^{\infty} dx \: e^{-x^4} \cos{(xt)} |_{large \: t} = 2^{1/6} \sqrt{\frac{\pi}{3}}\:  \frac{1}{t^{\frac{1}{3}}} e^{- \frac{2^{1/3} 3}{16} t^{4/3}} \cos{ \bigg(
  3^{\frac{3}{2}} \frac{2^{1/3}}{16} t^{4/3} - \frac{\pi}{6} \bigg) }  \nonumber \\
  \int_0^{\infty} dx \: e^{-x^4} \cos{(xt)} |_{small \: t} = 0.906402 - 0.153177 t^2 \approx 0.906402 e^{- 0.168995 t^2}
 \end{eqnarray}
 Then we rescale the integral in Equation (28) (using $B=\frac{1}{R^2 \:N a^4}$ and also ${\cal C} = \frac{B^{1/4}}{0.906402}$)
 \begin{eqnarray}
 \int_0^{\infty} dx \: e^{-B x^4} \cos{(xt)} = \frac{1}{B^{1/4}} \int_0^{\infty} dy \: e^{-y^4} \cos{(y \frac{t}{B^{1/4}})}
 \end{eqnarray}
 which means the short-time dependence of $q_1$ is
 \begin{eqnarray}
 q_1(t)|_{ST} \approx e^{- 0.168995 \: R \: \sqrt{N} a^2 t^2}
 \end{eqnarray}
i.e., the parameter $q_1$, the amplitude of the ``unmeasured'' state, decays with time like a Gaussian, with time scale $\tau^2_{ST}=\frac{1}{R \sqrt{N} a^2}$ in natural units. The dependence upon $R$ (the macroscopic number of possible end-states) and $a^2$ is identical to the results in \cite{Balian2}, with a very different physical realization. To reiterate, the decay of $q_1$ to $0$ represents the ``collapse'' of the superposed state into the other states where the electron is localized to slit $1$.

For the long-time limit, we use the small eigenvalue limit for the squared weights to obtain (with $A = \frac{1}{R \sqrt{N}}$) and also take the upper limit of the rapidly convergent integral off to infinity, to get
 \begin{eqnarray}
 q_1(t)|_{LT} \approx \int_0^{\infty} d \Omega \: \:  \frac{{\cal D}(\Omega)}{1+A^2 (\frac{\Omega^2}{a^2}-1)^2}  \cos {\Omega t} \approx {\cal C} \int_0^{\infty} d \Omega \: \:  \frac{1}{1+A^2 (\frac{\Omega^2}{a^2}-1)^2}  \cos {\Omega t}
 \end{eqnarray}
We use the relation (for small $A$),
 \begin{eqnarray}
 \int_0^{\infty} \frac{dx}{1 + A^2 (x^2-1)^2} \cos{x t} = \frac{ \pi  e^{- \frac{t}{\sqrt{2 A}} }}{2  \sqrt{A}}  \cos{  \left(   \frac{t}{\sqrt{2A}}-\frac{\pi}{4} \right) }
 \end{eqnarray}
 so the envelope of the graph of the parameter $q_1$ is
 \begin{eqnarray}
| q_1(t)|_{LT}| \approx {\cal C} \pi \frac{e^{- \frac{t}{\sqrt{2A}} }}{2  \sqrt{A}}
 \end{eqnarray}
 and obtain (upon suitably substituting for $A$)
 \begin{eqnarray}
 | q_1(t)|_{LT}| \sim  e^{- \sqrt{R \sqrt{N}} \:  a \:  t}
 \end{eqnarray}
 The parameter $q_1(t)$ falls off exponentially in the large $t$ limit. In addition, the collapse occurs over the time-scale $\tau_{LT}= \frac{1}{\sqrt{R \sqrt{N}}}$; this is unlike the situation in \cite{Balian2}, undoubtedly owing to the difference in the process of measurement and the manner (in that calculation) of the  connection of the phonon heat bath to the magnetic system. Here, in both cases, the scale of the time dependence includes the factor $R$, which is the macroscopic number of states accessible to the photons in the photo-multiplier. In the limit $R \rightarrow \infty$, we get instantaneous ``collapse'' of the amplitude to 0.
 
 Contrary to the analysis in \cite{Balian1}, we make no distinction between the ``registration'' of the ``collapsed'' wave-function and the ``collapse''. In our case, the ``collapse'' is simultaneous with ``registration'' - a macroscopic number of photons is produced in a continuum of states with corresponding localization of the electron to slit $1$.
 
 The analysis in \cite{Balian1, Balian2} also relies on using a system in a metastable state near a critical point to make a measurement. The practical problem with this is that such a system would settle into one or the other stable state (all up or all down spin) even with statistical fluctuation. The interaction hence needs to be proportional to $R$ right at the start. That is not our approach, as we see the number of modes appear in a rather natural way as the number of accessible channels for cascade photons.

\section{Connection to some general theoretical results}

In a series of interesting papers, Linden {\it et al}\cite{Popescu} have constructed arguments that collapse is generic in a variety of approaches to the problem of thermalization. The arguments are extremely persuasive and depend on some very general assumptions. One, that they need and we agree is not substantially important, is that the Hamiltonians have no degeneracies. That is not a major drawback in their analysis, since (as they point out), a small random perturbation will remove any degeneracies anyway. However, it is useful to note that nothing in our arguments above needs any assumptions about degeneracies. All we need is a diversity of states and eigenvalues (as described by Gershgorin's theorem) and the dependence of all  eigenvalues upon the first one (as in  Equation (21)).

Another point to note, as a difference to the thermalization literature, is that there is usually only one final equilibrium state, or at least one with static time-averaged properties. In contrast, a measurement process can lead to one of several results. It is not clear if their analysis can be extended to several possible equilibria, with no  dependence on initial conditions.

 \section{Recurrence times}
 
 In the general case, where the elements of $\cal G$ are random, with the constraints that it is a symmetric matrix of the form in Equation (12), the eigenvalues are likely distributed with the Wigner semi-circle distribution \cite{Wigner1} and the smallest separation is $\sim \frac{1}{R}$. The recurrence time for such a collection of coupled harmonic oscillators has been well-studied \cite{Hemmer, Venuti} and the results turn out to depend on three things
 \begin{enumerate}
 \item how exactly one defines ``closeness' to the initial state. In our example above, this is pretty clear, {\it viz.}  we need to get back to an uncollapsed state, which is quite unique in the Hilbert space. In the usual problem that is studied, that of coupled harmonic oscillators, it is defined by the phase angle difference between various oscillators $\Delta \phi$. In our case, this difference tends to zero $\sim \frac{1}{R}$, since a very small fraction of states could be considered uncollapsed.
 \item the separation between the frequencies of the oscillating states, which is obviously dominated by the smallest separation ($\sim \frac{1}{R}$) mentioned above and
 \item the number of degrees of freedom (number of oscillators), which is here the number of possible states accessible to the system. This number is the macroscopic $R$ used in the above.
 \end{enumerate}
 Under these conditions, the recurrence time for a collection of oscillators, which is fundamentally the problem we are studying here, is proportional to $\frac{1}{(\Delta \phi)^R}$, which is indeed {\it {\underline {exponential}}} in $R$. In fact, for a small system with 10 oscillators, as studied by Hemmer {\it et al}, this time is $\sim 10^{10}$ years.
 
 This argument is intended to buttress the points made in the introduction, that the long recurrence times help pick which state the system collapses into.
 
\section{Connections to Mott's thought experiment}
 
Mott \cite{Mott} considered the theoretical explanation of why $\alpha$-particle decay into an $s$-wave state leads to tracks in a cloud chamber, rather than a spherically symmetric fuzz around the decay site. Employing a simple argument, he showed that the various possible exit states of the $\alpha$-particle should be straight rays emanating from the decay site. In our language, each such ray would be equivalent to a ``detection at a slit''. From the above argument, there is an amplitude that any one of these rays is picked due to the interaction parameter with each of the macroscopic number of atoms in the cloud chamber. As described, once one of these rays picks up a small excess weight amongst the accessible states in Hilbert space, it is inevitable that it will ``win'' and the wave-function will collapse into that ray. All this is consistent with the analysis we have made in this paper.
 
  \section{Connections to Wigner's friend's experiment}
  
Wigner\cite{Wigner} discusses an experiment, that has been studied in various forms by others, including prominently by Deutsch \cite{Deutsch}. The experiment concerns a measurement made inside a closed system by a macroscopic observer (which could be an observing instrument, or Wigner's friend), as well as the observer outside viewing this closed system as a quantum system. The conclusion one is led to from the analysis in this paper is that if one accounts properly for the interaction Hamiltonian of the observer inside the closed system with all the parts of the system, one is led inexorably to the result that once a macroscopic number of states have flipped to the ``observed'' status, it is extremely unlikely that they will flip back {\it in toto}. Wigner (sitting outside) might not know this, but his lack of knowledge of the complete Hamiltonian cannot be regarded as a reasonable explanation for him expecting the system to stay in a superposed state. It would be akin to someone that doesn't watch a football game, but decides that it must not be decided simply because he did not watch it.

\begin{enumerate}
\item If Wigner and Wigner's friend both understand the interaction Hamiltonian between the measuring apparatus and the measured system, they will both agree that a random process will determine which outcome is selected, but invariably, one outcome will "win". This is consistent with our prior argument in the introduction, as well as through what we have shown in the analysis).
\item Once a particular outcome has "won", Wigner's friend knows (since she has looked at her measuring apparatus), but Wigner doesn't know the outcome.
\item As far as Wigner is concerned, the system is in a superposition of states. As far as the system and measuring apparatus is concerned, it is also in a superposition of states, however, one quantum fluctuation has consumed most of the particles and the corresponding measurement has "won". However, nothing discontinuous or outside the Schrödinger's equation has occurred. This is totally consistent with our analysis.
\item The fact that fluctuations occur is not inconsistent with time-evolution with a  deterministic Schrödinger's equation.
\item The fluctuation will inevitably reverse itself, however, the time scale will be of the order of $\sim e^M$, where $M$ is the number of particles involved. This might be much larger than multiple lifetimes of the universe. This is in line with the discussion on recurrence times in Section IV.
\item In a double slit experiment, suppose Wigner's friend knows which slit the electron went through, an outcome that was selected randomly by nature. If Wigner now follows the further path of the electron, he would find it going one way or the other, which is still consistent with an outcome selected randomly by nature. So there is no way, with a simple analysis of the outgoing particles only, to realize that the wave-function has collapsed.
\item Suppose, Wigner contrived to perform an interference experiment again, following his friend's double-slit set-up as follows. His friend sends out of her laboratory only those electrons whose ``which-path'' information she knows. Now, Wigner observes the density of arriving electrons on the screen - he will see no interference pattern, consistent with a wave-function that has collapsed - the ``which-path'' information is known to {\it some apparatus} known to his friend, even if that is not known to Wigner. This is consistent with the Delayed Choice experiment - an experiment performed on electrons for whom ``which-path'' information is available in some macroscopic form already will {\underline {not}} show interference\cite{Scully}. This is indeed  consistent with the notion discussed in Allahverdyan {\it et al} \cite{Balian5}
\end{enumerate}

 \section{Conclusions}

We have studied the measurement, using a macroscopic apparatus, of the double-slit experiment and deduced, from Schrödinger's equation, that the wave-function collapse upon measurement is a straightforward result of the interaction Hamiltonian. The time scales for the short-term and long-term collapse are $\tau_{ST}^2=\frac{1}{R \sqrt{N} a^2}$ and $\tau_{LT}= \frac{1}{\sqrt{R \sqrt{N}}}$ respectively. In addition, the short-term collapse follows a Gaussian, while the long-term behavior is exponential. This approach therefore connects the quadratic time dependence of the transition probability from Fermi's golden rule to the exponential behavior seen in radioactive decay, for instance. The calculation thus produces results parallel to \cite{Balian2}, except that the ``registration'' process is much quicker here (related to $\tau_{LT} \sim \frac{1}{\sqrt{R}}$) and doesn't behave as in the phonon bath in that reference. Some useful connections are made to the Wigner's friend and Mott thought experiments.

\section{Data Availability Statement}
Data sharing is not applicable to this article as no new data were created or analyzed in this study.

\section{Acknowledgments}
Useful discussions with Scott Thomas, Dan Friedan and Thomas Banks are acknowledged and especially to the latter for pointing out the similarities to Allahverdyan {\it et al}\: \cite{Balian1, Balian2, Balian3}. The hospitality of the NHETC and the Department of Astronomy and Physics at Rutgers University are sincerely acknowledged. Extremely useful comments from anonymous referees are acknowledged gratefully.


\section{Appendix I}

Consider the geometry shown in Fig. 14.
\begin{figure}[ht!]
\caption{Geometry of the Double Slit Experiment}
\centering
\includegraphics[scale=.9]{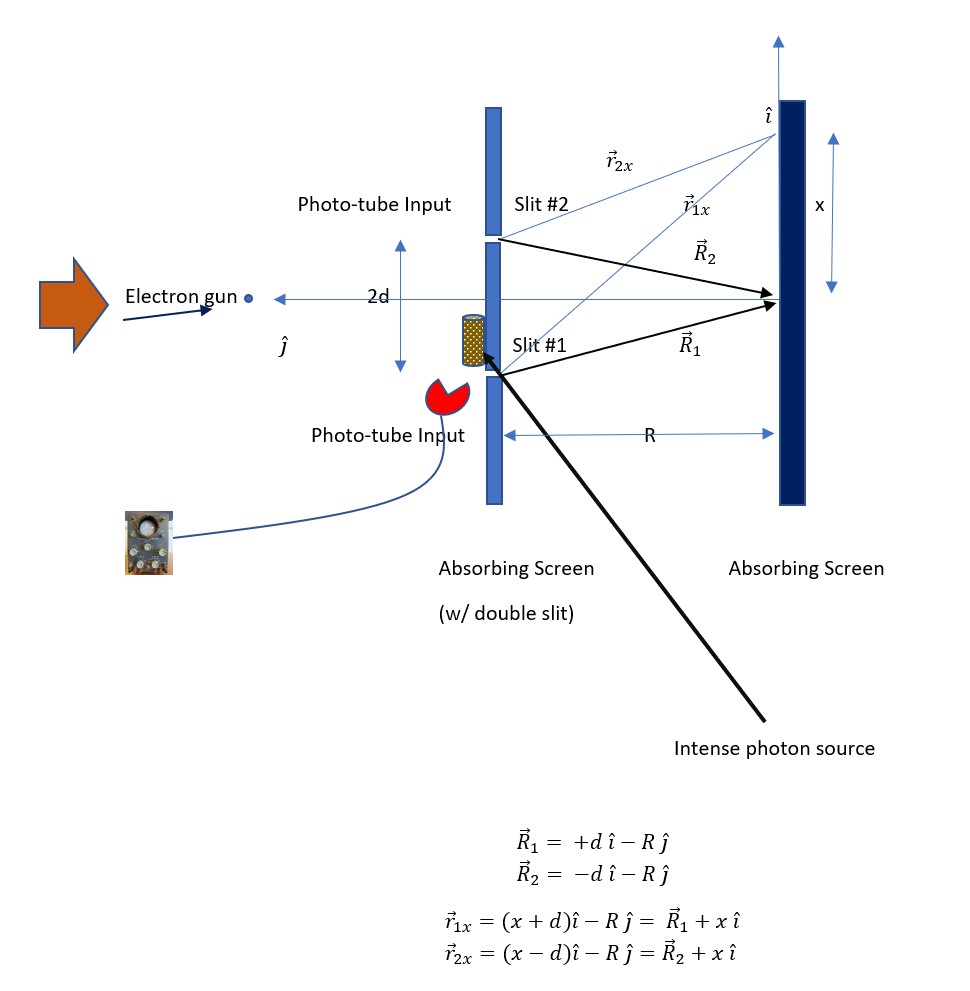}
\end{figure}

We can compute the wave-function at the point $x$ above the mid-point between the slits. We can think of the slits as the sources of spherical waves and write the wave-function at $x$ as the sum of spherical waves from each slit, {\it i.e.,}
\begin{eqnarray}
\psi_1(x) = \frac{e^{i k r_{1x}}}{\sqrt{4 \pi} \: r_{1x}^2} \nonumber \\
\psi_2(x) = \frac{e^{i k r_{2x}}}{\sqrt{4 \pi} \: r_{2x}^2} \nonumber \\
\psi(x) = \frac{\psi_1(x)+\psi_2(x)}{\sqrt{2}}
\end{eqnarray}
where $r_{1x}, r_{2x}$ are as defined in Fig. 14. The $k$ is defined from the electron's momentum $\hbar k$. Note that we have two interesting limits, {\it i.e., } $d \ll R \sim x$ and $d \ll R \ll x$.
We write (see behavior in Fig. 15)
\begin{eqnarray}
r_{1x}-r_{2x} \equiv \Phi = 
\begin{cases}
    \frac{2 d x}{R}& \text{if } d \ll R \sim x \\
    2 d              & \text{if } d \ll R \ll x
\end{cases}
\end{eqnarray}
With this, we compute
\begin{eqnarray}
P(x) = |\psi(x)|^2 = \nonumber \\
\frac{2}{(R^2+d^2)^2} \frac{1}{(1+\frac{x^2}{R^2+d^2})^2 - \frac{4 d^2 x^2}{(R^2 + d^2)^2}} \: \bigg[ \frac{(1+\frac{x^2}{R^2+d^2})^2 + \frac{4 d^2 x^2}{(R^2 + d^2)^2}}{(1+\frac{x^2}{R^2+d^2})^2 - \frac{4 d^2 x^2}{(R^2 + d^2)^2}}  + \cos \Phi \bigg]
\end{eqnarray}
\begin{figure}[ht!]
\caption{$\Phi \: vs. \: x$ for $d=1, R=10$}
\centering
\includegraphics[scale=.9]{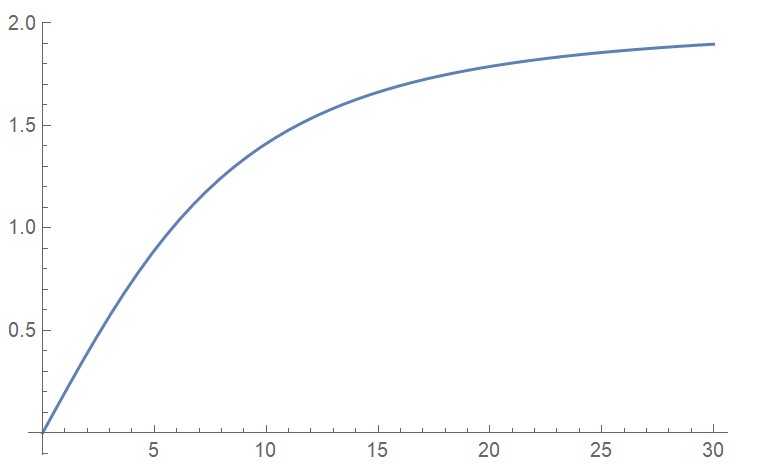}
\end{figure}

We have the limiting forms
\begin{eqnarray}
P(x) = 
\begin{cases}
    \frac{4 \cos^2 \frac{\Phi}{2}}{R^4}& \text{if } d \ll R \sim x \\
     \frac{4 \cos^2 \frac{\Phi}{2}}{x^4}              & \text{if } d \ll R \ll x
\end{cases}
\end{eqnarray}
which we plot (as an example for, additionally $k=20 \pi$)

\begin{figure}[ht!]
\caption{$P(x) \: vs. \: x$ for $d=1, R=10, k=20 \pi$}
\centering
\includegraphics[scale=.9]{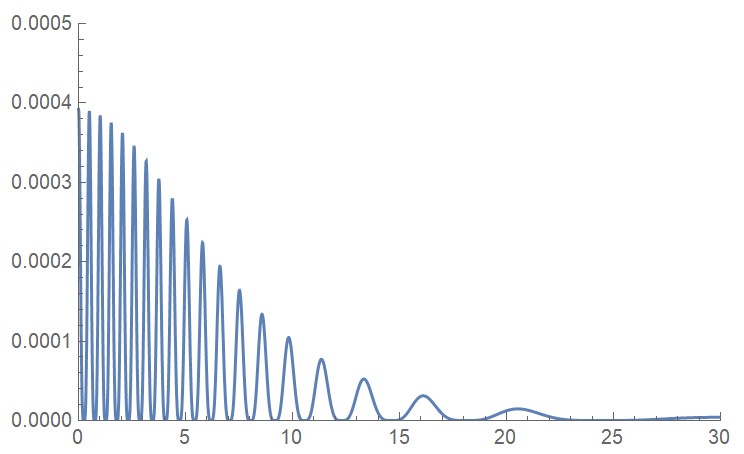}
\end{figure}

\section{Appendix II}

A matrix ${\cal G}$ is positive semi-definite if $x^T {\cal G} x \ge 0$ for any vector $x$.
Consider the matrix defined as 
\begin{eqnarray}
{\cal G} = \left(\begin{array}{cccccccccc} a_1^2 & \: \: \:  0 &\: \: \:  a_1 a_2 & \: \: \: 0 \: \: \: &  \: \: \: 0 \: \: \: & ... & ... & ... &... &\: \: \: 0 \: \: \: \\
					     0 & \: \: \:  a_1^2+a_2^2 & \: \: \: 0 &  \: \: \: a_2 a_3 \: \: \: &\: \: \:  0 &  ... & ... & ... &... &0 \\
					     a_1 a_2 & \: \: \: 0 & \: \: \: a_2^2+a_3^2 & \: \: \: 0 \: \: \: &a_3 a_4 &   ... & ... &... &... &0 \\
					     0 & \: \: \: a_2 a_3 & \: \: \: 0 & \: \: \: a_3^2+a_4^2 \: \: \: &\: \: \: 0 &   ... & ... &... &... &0 \\
					     0 & \: \: \: 0 & \: \: \: a_3 a_4 & \: \: \: 0 \: \: \:   &\: \: \: a_4^2+a_5^2 &   ... & ... &... &... &0 \\
					     0 & \: \: \: 0 & \: \: \: 0 & \: \: \: a_4 a_5 \: \: \: &\: \: \: 0 & ... & ... &   ... &... &0 \\
					     0 & \: \: \: 0 & \: \: \: 0 & \: \: \: 0 \: \: \: &\: \: \: a_5 a_6 & ... & ... & ... & ... &0 \\
					     0 & \: \: \: 0 & \: \: \: 0 & \: \: \: 0 \: \: \: &\: \: \: ... & ... & ... & ... & ... &0 \\
					     0 & \: \: \: ... & \: \: \: ... & \: \: \: ... \: \: \: &\: \: \: ... & ... & ... & ... & ... &0 \\
					     0 & \: \: \: ... & \: \: \: ... & \: \: \: ... \: \: \: &\: \: \: ... & ... & ... & ... & ... &0 \\
					     0 & \: \: \: ... & \: \: \: ... & \: \: \: ... \: \: \: &\: \: \: ... & ... & ... & ... & ... &a_{Q-2} a_{Q-1} \\
					     0 & \: \: \: ... & \: \: \: ... & \: \: \: ... \: \: \: &\: \: \: ... & ... & ... & ... & ... &0 \\
					     0 & \: \: \: 0 & \: \: \: 0 & \: \: \: 0 \: \: \: &\: \: \: 0 & ... & ... & ... & ... &a_{Q-1}^2 
					      \end{array}\right)
\end{eqnarray}
For an arbitrary vector $x$, with components $x_1, x_2, ..., x_Q$ and the definitions $a_{-1}=a_{0}=a_{Q}=a_{Q+1}=0$ and $x_{-1}=x_0=x_{Q+1}=x_{Q+2}=0$, we write the scalar as
\begin{eqnarray}
\left(\begin{array}{cccc} x_1 & x_2 &\: \: \:  . & x_Q
					      \end{array}\right)
					      \left(\begin{array}{cccccccccc} a_0^2+a_1^2 & \: \: \:  0 &\: \: \:  a_1 a_2 & \: \: \: 0 \: \: \: &  \: \: \: 0 \: \: \: & ... & ... & ... &... &\: \: \: 0 \: \: \: \\
					     0 & \: \: \:  a_1^2+a_2^2 & \: \: \: 0 &  \: \: \: a_2 a_3 \: \: \: &\: \: \:  0 &  ... & ... & ... &... &0 \\
					     a_1 a_2 & \: \: \: 0 & \: \: \: a_2^2+a_3^2 & \: \: \: 0 \: \: \: &a_3 a_4 &   ... & ... &... &... &0 \\
					     0 & \: \: \: a_2 a_3 & \: \: \: 0 & \: \: \: a_3^2+a_4^2 \: \: \: &\: \: \: 0 &   ... & ... &... &... &0 \\
					     0 & \: \: \: 0 & \: \: \: a_3 a_4 & \: \: \: 0 \: \: \:   &\: \: \: a_4^2+a_5^2 &   ... & ... &... &... &0 \\
					     0 & \: \: \: 0 & \: \: \: 0 & \: \: \: a_4 a_5 \: \: \: &\: \: \: 0 & ... & ... &   ... &... &0 \\
					     0 & \: \: \: 0 & \: \: \: 0 & \: \: \: 0 \: \: \: &\: \: \: a_5 a_6 & ... & ... & ... & ... &0 \\
					     0 & \: \: \: 0 & \: \: \: 0 & \: \: \: 0 \: \: \: &\: \: \: ... & ... & ... & ... & ... &0 \\
					     0 & \: \: \: ... & \: \: \: ... & \: \: \: ... \: \: \: &\: \: \: ... & ... & ... & ... & ... &0 \\
					     0 & \: \: \: ... & \: \: \: ... & \: \: \: ... \: \: \: &\: \: \: ... & ... & ... & ... & ... &0 \\
					     0 & \: \: \: ... & \: \: \: ... & \: \: \: ... \: \: \: &\: \: \: ... & ... & ... & ... & ... &a_{Q-2} a_{Q-1} \\
					     0 & \: \: \: ... & \: \: \: ... & \: \: \: ... \: \: \: &\: \: \: ... & ... & ... & ... & ... &0 \\
					     0 & \: \: \: 0 & \: \: \: 0 & \: \: \: 0 \: \: \: &\: \: \: 0 & ... & ... & ... & ... &a_{Q-1}^2 + a_Q^2 
					      \end{array}\right) \left(\begin{array}{c} x_1 \\
					      	x_2 \\
					      	.\\
					      	.\\
					      	.\\
					      	.\\
					      	.\\
					      	.\\
					      	.\\
					      	.\\
					      	.\\
					      	.\\
					      	x_Q
					      \end{array}\right) \nonumber
\end{eqnarray}
Hence
\begin{eqnarray}
x^T {\cal G} x = \bigg( a_{-1} a_0 x_{-1} x_1 + (a_0^2+a_1^2) x_1^2 + a_1 a_2 x_1 x_3 \: \: \: \: \: \: \: \: \: \: \: \: \: \: \: \: \: \: \: \: \: \: \: \: \: \: \: \: \: \: \: \: \: \: \: \: \: \: \: \: \: \: \: \: \: \: \: \: \: \: \: \: \: \: \: \: \: \: \: \: \nonumber \\
+ \: \: a_{0} a_1 x_{0} x_2 + (a_1^2+a_2^2) x_2^2 + a_2 a_3 x_2 x_4 \: \: \: \: \: \: \: \: \: \: \: \: \: \: \: \: \: \: \: \: \: \: \: \: \: \: \: \: \: \: \: \: \: \: \: \: \: \: \: \: \: \: \: \: \: \: \: \: \: \: \: \: \: \: \: \: \: \: \: \: \: \: \: \: \: \nonumber \\
+ \: \: a_{1} a_2 x_1 x_3 + (a_2^2+a_3^2) x_3^2 + a_3 a_4 x_3 x_5 \: \: \: \: \: \: \: \: \: \: \: \:  \: \: \: \: \: \: \: \: \: \: \: \: \: \: \: \: \: \: \: \: \: \: \: \: \: \: \: \: \: \: \: \: \: \: \: \: \: \: \: \: \: \: \: \: \: \: \: \: \: \: \: \: \: \nonumber \\
+ \: ...  \: \: \: \: \: \: \: \: \: \: \: \: \: \: \: \: \: \: \: \: \: \: \: \: \: \: \: \: \: \: \: \: \: \: \: \: \: \: \: \: \: \: \: \: \: \: \: \: \: \: \: \: \: \: \: \: \: \: \: \: \: \: \: \: \: \: \: \: \: \: \: \: \: \: \: \: \: \: \: \: \: \: \: \: \: \: \: \: \: \:\: \: \: \: \: \: \: \: \: \: \: \:  \nonumber \\
+ \: \: a_{Q-2} a_{Q-1} x_{Q-2} x_Q + (a_{Q-1}0^2+a_Q^2) x_1^2 + a_Q a_{Q+1} x_Q x_{Q+2} \bigg) \: \: \: \: \: \: \: \: \: \: \: \: \: \: \: \: \: \: \: \: \: \: \: \:  \nonumber \\
= a_1^2 x_2^2 + a_{Q-1}^2 x_Q^2 + (a_1 x_1+a_2 x_3)^2 + (a_2 x_2+a_3 x_4)^2 + ... + (a_{Q-2} x_{Q-2}+a_{Q-1} x_Q)^2 \ge 0 \nonumber 
\end{eqnarray}
and is true $\forall x$. The matrix is, hence, positive-definite.

\section {Appendix III}

In order to produce a microscopic basis for Equation (3), one should look at an elementary process in a very localized volume near a slit, from the following term in the QED action
\begin{eqnarray}
{\cal A}_{int} = - ie \: \int d^4 x  \:  \: {\bar \psi} \: \gamma^{\mu} A^{\mu} \psi
\end{eqnarray}

Consider the physics of the situation with the electron. Initially it is delocalized between the slits; post interaction with the photomultiplier's field, it gets localized to the slit $1$. The dimensions of the slit is usually compared to the electron's Compton wavelength (see Jonsson {\it al}).

Hence, considering Equation (3), the process is that electron comes into a region, with energy $\omega_e$, leaves with energy $\omega_e$ and also causes the emission of a photon of energy $\omega_p$. Some energy was added in the process. The energy addition must come through the interaction of the electron with a background field. The background field could be considered very concentrated in the region around slit $1$. We also note that for the geometries considered here, $\omega_p \ll \omega_e$, as we only need to localize the electron to within a $\mu$ or so with photons, while the electron wave's wavelength is much shorter.

To this end, we consider a second-order process, with a very intense (localized in space around the slit $1$) auxiliary field $A_{\mu}^{classical}$ which is large but constant in the region of slit $1$ and also constant in time. We then get the approximate second-order term below (as well as its hermitian conjugate), with the usual annihilation/creation operators for photons ($a, a^{\dagger}$) and electrons ($c, c^{\dagger}$), and time averaged over 
\begin{eqnarray}
= - e^2  \Delta V \Delta t  \:  \: {\bar \psi}(\vec x_1) {\slashed A} \psi(\vec x_1 ) \: \: < {\bar \psi}(\vec x_1)  {\slashed A}^{classical} \psi(\vec x_1 ) >   \nonumber \\
= (constants) \: c^{\dagger}(\vec x_1) e^{-i \omega_{e} t} \: \: \times  \: \:  \: \: \: a^{\dagger} e^{-i \omega_p t} \:  \: \: \times  \: \: \: c(\vec x_1)  e^{ i \omega_{e} t} \nonumber \\
\approx \Gamma_1 e^{-i \omega_p t} a^{\dagger} c^{\dagger}_1 c_1 \: \: \: \:  \: \: \: \:  \: \: \: \:  \: \: \: \:  \: \: \: \:  \: \: \: \:  \: \: \: \:  \: \: \: \:  \: \: \: \:  \: \: \: \: 
\end{eqnarray}
which is what we used (along with its hermitian conjugate) in the rather idealized calculation (Equation 3).

It is clear from the above that energy is being added by the background field.

\begin{figure}[ht!]
\caption{Electron Scattering Off an background field and producing a photon - here $\omega_{\gamma} = \omega_p$}
\centering
\includegraphics[scale=.8]{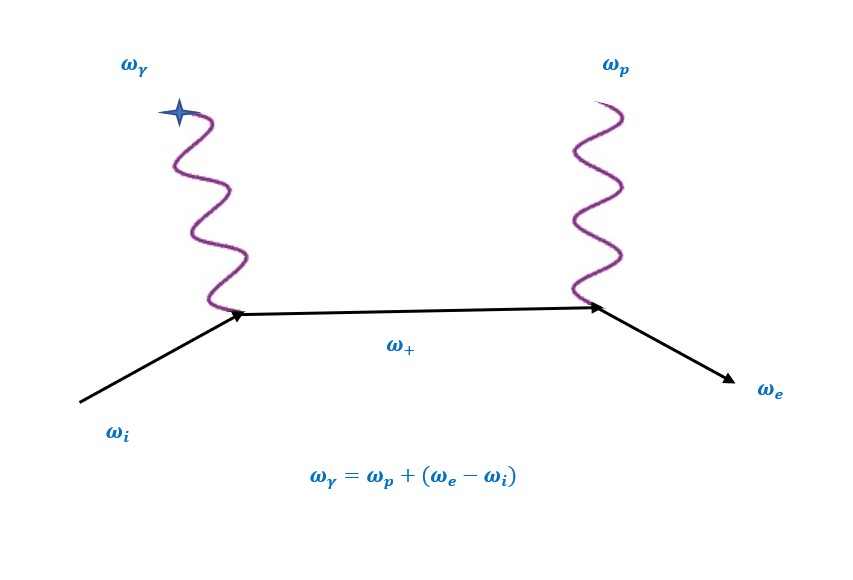}
\end{figure}


\section{References}
\begin{thebibliography}{25}


\bibitem{Griffiths} Griffiths,  R . J. \textit{Consistent Histories and the Interpretation of Quantum Mechanics} J. Stat. Phys. 36, 219 (1984).
\bibitem{Omnes}  Omn`es, R . \textit{Consistent Interpretations of Quantum Mechanics}  Rev. Mod. Phys. 64, 339 (1992).
\bibitem{Hartle1} Gell-Mann, M . \& Hartle, J. B. \textit{Alternative Decohering Histories in Quantum Mechanics}, Caltech preprint CALT-68-1694  (1991)
\bibitem{Hartle2} Gell-Mann,  M .  \& Hartle, J. B. \textit{Quantum Mechanics in the Light of Quantum Cosmology} p. 425 in Complexity, Entropy, and the Physics of Information,
W. H. Zurek, editor (Addison-Wesley) (1990)
\bibitem{Everett} Everett, H. \textit{`Relative state' formulation of quantum mechanics} Reviews of Modern Physics. 29 (3): 454–462 (1957).
\bibitem{Maximilian} Scholosshauer, Maximilian \textit{Quantum Decoherence} arxiv: 9111.066282v1
\bibitem{Zeh} Zeh, H . D. \textit{On the Irreversibility of Time and Observation in Quantum Theory}, in Enrico Fermi School of Physics IL, B. d’Espagnat, ed. (Academic Press, New York) (1971).
\bibitem{Zurek} Zurek,  W . H. \textit{Pointer Basis of Quantum Apparatus: Into What Mixture Does the Wave Packet Collapse?}, Phys. Rev. D 24, 1516  (1981)
\bibitem{Weisskopf} Weisskopf V. \&  Wigner E., Z. Phys. 63, 54 (1930)
\bibitem{Weisskopf} Weisskopf V. \&  Wigner E., Z. Phys. 65, 18 (1930)
\bibitem{Balian1} Allahverdyan, Armen E. ; Balian, Roger \& NieuwenHuizen, Theo M.  \textit{The Quantum Measurement Process: Lessons from an exactly solvable model}', arxiv:quant-ph/0702135v2 (2007)
\bibitem{Sorgel} Sörgel,  L., Hornberger, K. \textit{Unraveling quantum Brownian motion: Pointer states and their classical trajectories} Phys. Rev. A 92 (2015) 062112.
\bibitem{Balian2} Balian, Roger \textit{Emergences in Quantum Measurement Processes}', KronoScope 13:1 85-95 (2013)
\bibitem{Balian3} Allahverdyan, Armen E. ; Balian, Roger \& NieuwenHuizen, Theo M. \textit{Curie-Weiss model of the quantum measurement process}, Europhysics Letters 61.4 452-458 (2003).
\bibitem{Balian5} Allahverdyan, Armen E. ; Balian, Roger \& Nieuwenhuizen,  Theo M. \textit{Understanding quantum measurement from the solution of dynamical models} 	Phys. Rep. 525 (2013) 1-166
\bibitem{Feynman} Feynman, R. P. \textit{Lectures in Physics, Vol. 3} Wiley Eastern (1967).
\bibitem{Coleman} Coleman, S.  \textit{Quantum mechanics in your face.} APS New England Sectional Meeting (1994).
\bibitem{Murayama} Murayama, H. \textit{Lecture Notes 221A} http://hitoshi.berkeley.edu/221A/coherentstate.pdf
\bibitem{Gersh} Gerschgorin, S.  \textit{Über die Abgrenzung der Eigenwerte einer Matrix}, Izv. Akad. Nauk. USSR Otd. Fiz.-Mat. Nauk (in German), 6: 749–754 (1931).
\bibitem{Boyd} Boyd, John P.   \textit{The Fourier Transform of the quartic Gaussian $e^{-Ax^4}$: Hypergeometric functions, power series, steepest descent asymptotics and hyperasymptotics and extensions to $e^{-Ax^{2n}}$}, Applied Mathematics and Computation 0096-3003 (2014). Also at http://dx.doi.org/10.1016/j.amc.2014.05.001 (2014).
\bibitem{Popescu} Linden, N., Popescu, S., Short, A. J., Winter, A. \textit{Quantum Mechanical evolution towards thermal equilibrium} Phys. Rev, E 79:061103 (2009)
\bibitem{Wigner1} Wigner, E. P.  \textit{On the Distribution of the Roots of Certain Symmetric Matrices.} Ann. of Math. 67, 325-328, 1958. 
\bibitem{Hemmer} Hemmer, P. C., Maximon, L. C., Wergeland, H. \textit{Recurrence Time of a Dynamical System} Phys. Rev. Vol. 111, 3 (1958)
\bibitem{Venuti} Venuti, L. C. \textit{The recurrence time in quantum mechanics} arxiv:1509.04352v2
\bibitem{Mott} Mott, N. F. , \textit{The wave mechanics of $\alpha$-ray tracks.} Proc. Roy. Soc. London. A (1929).
\bibitem{Wigner} Wigner, E. P. \textit{Remarks on the mind-body question} ``The Scientist Speculates'', I. J Good (eds), Heinemann (1961)
\bibitem{Deutsch} Deustch, D. \textit{Quantum theory as a universal physical theory} Intl. J. of Theoretical Physics, {\bf 24}(1):1-41 (1985) 
\bibitem{Scully} Scully, M. O., Druhl, K. \textit{Quantum eraser: A proposed photon correlation experiment concerning observation and "delayed choice" in quantum mechanics} Phys. Rev. A. 25(4) (1982)
\bibitem{Borg} Borg, Frank  \textit{Quantum Profiles and Paradoxes}', arxiv.physics 0611164v1 16-Nov-2006 (1962).
\bibitem{Zwanzig} Zwanzig, R. \textit{Ensemble method in the theory of irreversibility}, J. Chem. Phys. 33 (1960) 1338-1341.
\bibitem{Vedral} Vedral, V. \textit{Observing the Observer} arxiv:1603.04583.
\bibitem{Balian4} Allahverdyan, Armen E. ; Balian, Roger \& Nieuwenhuizen,  Theo M. \textit{A sub-ensemble theory of ideal quantum measurement processes} Annals of Physics 376 (2017), 324-352
\bibitem{Jonsson} Jönsson, C. \textit{Electron diffraction at multiple slits} American Journal of Physics 42 4-11 (1974 )
\bibitem{Jonsson2} Jönsson, C. \textit{Elektroneninterferenzen an mehren kunstlich hergestellten Feinspalten} Zeitschrift fur Physik 161 (4), 454-474
\bibitem{Balian6}  ``dephasing'' refers to the phenomenon that the states lose phase coherence; then the phases get randomized in interaction with a bath of other oscillators, which is referred to as ``decoherence'' . See Ref. 35 for a lucid description.



\end{thebibliography}
\end{document}